\documentstyle[aps,epsf]{revtex}
\newcommand{\bm}{\mathbf}

\newcommand{\oee}{\mathop{\varepsilon}\nolimits}

\newcommand{\sr}{\phantom{0}}
\begin{document}
\draft
\bibliographystyle{revtex}
\title{Evaluation of the low-lying energy levels of two- and
       three-electron configurations for multi-charged ions}
\author{O.~Yu.~Andreev\footnote{Email: oleg@landau.phys.spbu.ru},${}^1$ 
L.~N.~Labzowsky,${}^1$
        G.~Plunien,${}^2$ and G.~Soff${}^2$}
\address{${}^1$ {Institute of Physics,
             St.~Petersburg State University, Ulyanovskaya ul. 1,
             198904,
             Petershof, St.~Petersburg, Russia}}
\address{${}^2$ {Institut f\"ur Theoretische Physik,
             Technische Universit\"at Dresden,
             Mommsenstra{\ss}e 13, D-01062, Dresden, Germany}}
\date{\today}
\maketitle
\begin{abstract}
Accurate QED evaluations of the one- and two-photon
interelectron interaction for low lying two- and
three-electron configurations
for ions with nuclear charge numbers
$60\le Z \le 93$
are performed.
The three-photon interaction is also partly taken into account.
The Coulomb gauge is employed.
The results are compared with available experimental data and
with different calculations.
A detailed investigation of the behaviour of the energy levels
of the configurations
$1s_{1/2}2s_{1/2} \,{}^1\! S_0$,
$1s_{1/2}2p_{1/2} \,{}^3\! P_0$
near the crossing points
$Z=64$ and $Z=92$ is carried out.
The crossing points are important for the future experimental
search for parity nonconserving (PNC) effects in highly
charged ions.
\end{abstract}
\section{Introduction}
\label{introduction}
During the most recent years the energy levels of two- and
three-electron configurations are under very intensive
experimental and theoretical investigation.
\par
Accurate calculations of the energy levels for
the two-electron configurations
$1s_{1/2}2s_{1/2} \,{}^1\! S_0$,
$1s_{1/2}2p_{1/2} \,{}^3\! P_0$,
and
$1s_{1/2}2s_{1/2} \,{}^3\! S_1$
were already performed in
\cite{drake88,plante94}.
In these papers the interelectron interaction
has been considered in various approximations:
on the basis of
variational Schr{\"o}dinger-wave functions
with inclusion of relativistic correlations
\cite{drake88}
and using relativistic many-body perturbation theory (RMBPT)
\cite{plante94}.
For a long time these approaches
defined the standard concerning the level of accuracy.
In recent years due to new developments in experimental
and theoretical methods
the  necessity to improve the accuracy of 
calculations
became urgent.
Recently, rigorous QED
evaluations of two-photon exchange corrections
for low lying configurations
\cite{mohr00,andreev01}
have been performed.
\par
First calculations of the energy levels for the three-electron
configuration 
have been presented
in
\cite{johnson88p2764,kim91,blundell93,chen95,persson96}.
As 
in the early papers on two-electron ions
in these calculations the two- and many-photon exchange 
has been
considered approximately.
Complete QED calculations of the two-photon exchange for
three-electron ions have been performed in
\cite{yerokhin01,andreev01}.
Within the framework of RMBPT
the three-photon exchange has also been taken
into account in
\cite{zherebtsov00,yerokhin01,andreev01}.
\par
In this paper we present an extension of previous calculations for 
two-and three-electron ions
\cite{andreev01}
for a variety of 
$Z$ values.
Here we include also the evaluations for the
$1s_{1/2}2p_{1/2} \,{}^3\! P_0$
level, which have not been performed in
\cite{andreev01}.
Special attention has been paid to eleborate the behaviour of the
$1s_{1/2}2s_{1/2} \,{}^1\! S_0$
and
$1s_{1/2}2p_{1/2} \,{}^3\! P_0$
levels near
$Z=64$
and
$Z=92$,
where they become
very close to each other.
Accordingly, these system  become
suitable for monitoring parity-nonconserving (PNC)
effects.
Experimental investigations of PNC effects in
two-electron highly
charged ions are under intensive discussion
\cite{schaeter89,karasiev92pla,dunford96,labzowsky01}.
The preparation of such experiments requires a precise knowledge of
the exact difference between these energy levels.
\section{Theory}
\label{theory}
\begin{figure}
\centerline{ \mbox{ \epsfxsize=0.1\textwidth \epsffile{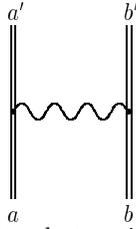} }}
 \caption{
Feynman graph, describing the first-order interelectron
interaction.
The double solid lines correspond to bound electrons in the field
of the nucleus, the wavy line corresponds to the sum of the
Coulomb and Breit (transverse) photons.
If
$a'=a$
and
$b'=b$
the graph is called ``direct'', in case
$a'=b$,
$b'=a$
we call it ``exchange'' graph.
The latter name should be understood in connection with respect to 
permutation
symmetry.
}
\label{fig1}
\end{figure}

\begin{figure}
\begin{center}
\end{center}
\centerline{ \mbox{ \epsfxsize=0.3\textwidth \epsffile{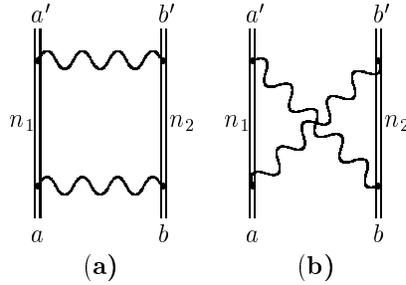} }}
\caption{
Feynman graphs describing the second-order interelectron interaction.
The graph a) is called ``box'' and the graph b) 
is called ``cross''.
Notations are the same as in Fig. 1.
By $n_1, n_2$
the summation over intermediate states is indicated.
}
\label{fig2}
\end{figure}

In this article we evaluate corrections to the energy levels
due to  photon exchange.
To calculate these corrections we employ the adiabatic
$S$-matrix
approach
\cite{gellmann51,sucher57}
and the line profile approach (LPA)
\cite{labzowsky93karasiev}.
Both methods are based on the Furry picture
\cite{furry51},
which describes the many-electron atom as a set of bound electrons,
moving in the external field of the nucleus and interacting with 
each other via the exchange of photons.
With the aid of 
the Feynman 
rules
for bound-state QED
\cite{labzowsky93b,labzowsky93only}
the processes giving rise to
corrections to the energy levels can be represented in terms of 
Feynman graphs.
The photon-exchange corrections evaluated in this article
are depicted in
Figs.\,\ref{fig1}-\ref{fig5}.
\par
As it has been shown in
Refs.\,\cite{labzowsky86b,labzowsky93b}
the $S$-matrix approach is best suited for the evaluation of
corrections described by the irreducible parts of the diagrams.
However, its application 
to the evaluation of the reducible parts becomes rather
complicated.
Accordingly, for calculating the reducible parts we utilized the LPA.
For a detailed description and for the application of this 
method we refer to
\cite{andreev01}.
\par
The Coulomb gauge will be used throughout this paper.
The photon propagators for Coulomb
(${\rm g}={\rm c}$)  
and transverse
(${\rm g}={\rm t}$)
photons are
given by
\cite{labzowsky93b,labzowsky93only}:
\begin{eqnarray}
D_{\mu_1 \mu_2}^{\rm g}(x_1,x_2)
\label{dfotone}
&=&
\frac{1}{2\pi i}
\int\nolimits_{-\infty}^{\infty}d\Omega\,
I_{\mu_1 \mu_2}^{\rm g}(\Omega,r_{12})\, e^{i\Omega(t_1-t_2)}
\end{eqnarray}
together with the temporal Fourier transforms
\begin{eqnarray}
\label{ic}
I_{\mu_1 \mu_2}^{\rm c}(\Omega,r_{12})
&=&
\frac{\delta_{\mu_1 4} \delta_{\mu_2 4}}{r_{12}}
\end{eqnarray}
and
\begin{eqnarray}
\label{it}
I_{\mu_1 \mu_2}^{\rm t}(\Omega,r_{12})
&=&
\left(\frac{\delta_{\mu_1 \mu_2}}{r_{12}} e^{i|\Omega|r_{12}}
+
\nabla_{1\mu_1}\nabla_{2\mu_2}\frac{1}{r_{12}}\,
\frac{1-e^{i|\Omega|r_{12}}}{|\Omega|^2}
\right)
(1-\delta_{\mu_1 4})(1-\delta_{\mu_2 4})
\,.
\end{eqnarray}
\subsection{The two-electron configurations}
\label{theorytwo}
At first we consider the photon exchange corrections for two-electron
configurations.
The wavefunction of a two-electron configurations is represented by
\begin{eqnarray}
\Psi_{JMj_1j_2l_1l_2}({\bm r}_{1}, {\bm r}_{2})
\label{wavef}
&=&
N
\sum\limits_{m_1m_2}
\mbox{\rm C}^{j_1j_2}_{JM}(m_1m_2)
\left[
\psi_{j_1l_1m_1}({\bm r}_{1})
\psi_{j_2l_2m_2}({\bm r}_{2})
-
\psi_{j_1l_1m_1}({\bm r}_{2})
\psi_{j_2l_2m_2}({\bm r}_{1})
\right]
\,,
\end{eqnarray}
where
$N=1/2$
for equivalent electrons and
$N=1/\sqrt{2}$
for non-equivalent electrons,
$\mbox{\rm C}^{j_1j_2}_{JM}(m_1m_2)$
is a Clebsch-Gordan symbol.
By means of
Eq. (\ref{wavef})
we can specify the
configuration
$1s2s\,{}^3\!S_1$
by setting 
$a,b=1s_+,2s_+$, 
where 
$\pm$ 
denote the two different projections of the total electron angular 
momentum
and we can derive
the energy corrections 
according to the formula:
\begin{eqnarray}
\label{h3s1}
\Delta E (1s2s\,{}^3\!S_1)
&=&
F_{1s_+2s_+;1s_+2s_+}
\,,
\end{eqnarray}
\begin{eqnarray}
F_{ab;cd}
\label{def2det}
&=&
F_{abcd}-F_{bacd}
\,.
\end{eqnarray}
Here
$F_{ab\ldots}$
denotes a function of one-electron states which are described by
wave functions
$\psi_{a}, \psi_{b},\ldots$.
The form of the function
$F$
depends on the type of 
Feynman graph 
under cosideration
(see below).
For the electron configurations
$1s2s\,{}^1\!S_0$
and
$1s2p\,{}^1\!P_0$
the energy corrections 
are given by
\begin{eqnarray}
\Delta E (1s2s\,{}^1\!S_0)
&=&
F_{1s_-2s_+;1s_-2s_+}
-
F_{1s_+2s_-;1s_-2s_+}
\,,
\end{eqnarray}
and
\begin{eqnarray}
\Delta E (1s2p\,{}^3\!P_0)
&=&
F_{1s_-2p_+;1s_-2p_+}
-
F_{1s_+2p_-;1s_-2p_+}
\, ,
\end{eqnarray}
respectively.
The corrections due to one-photon exchange are represented
by the graph in
Fig.\,\ref{fig1}.
This diagram is irreducible so
that the
$S$-matrix
approach can be applied 
leading to
\begin{eqnarray}
F_{a'b'ab}^{(1)}
\label{ant}
&=&
\mathop{{\sum}}_{\rm g}
I^{\rm g}(\oee_{a'}-\oee_a)_{a'b'ab}
\,.
\end{eqnarray}
Here we have introduced the following notation
(see definitions Eqs. 
(\ref{ic}) and (\ref{it}))
\begin{eqnarray}
I_{a'b'ab}^{\rm g}(\Omega)
&\equiv&
\label{defmat}
\sum_{\mu_1\mu_2}
\int \overline\psi_{a'}({\bm r}_1)\overline\psi_{b'}({\bm r}_2)
\gamma_{\mu_1}^{(1)} \gamma_{\mu_2}^{(2)}
I_{\mu_1 \mu_2}^{\rm g}(\Omega,r_{12})
\psi_{a}({\bm r}_1)\psi_{b}({\bm r}_2)\,d^3r_1 d^3r_2 
\,,
\end{eqnarray}
where the Dirac matrices
$\gamma_{\mu_i}^{(i)}$
are acting on wave functions depending on spatial variables
${\bm r}_i$, 
respectively.
For
${\rm g = c}$
Eq. (\ref{ant})
determines the first-order Coulomb correction while for
${\rm g = t}$
we obtain the first-order Breit correction.
\par
The two-photon exchange corrections are represented by the graphs in
Fig.\,\ref{fig2}.
The ``box'' diagram is reducible.
Its reducible part is defined by the condition
$\oee_{n_1}+\oee_{n_2}=\oee_{a}+\oee_{b}$.
The ``cross'' diagram is irreducible.
However, it is most convenient
to extract the contribution with
$n_1$, $n_2$
equal to
$a$ or $b$
and to treat it like
a reducible part of the ``cross'' diagram.
Contributions due to states
$n_1$, $n_2$
included in the reducible parts are called reference
state contribution.
Application of the $S$-matrix approach for calculating the irreducible
part and of the LPA for the reducible part of the ``box'' and ``cross''
diagrams, respectively,
results in
the explicit formulas:
\begin{eqnarray}
F_{a'b'ab}^{(2)(\rm box,irr)}
\label{boxirr}
&=&
\mathop{{\sum}}_{\rm g g'}
\mathop{{\sum}'}_{n_1 n_2}
\left\{ \vphantom{\frac{8}{8}}\right.
\frac{i}{2\pi}\int\limits_{-\infty}^{\infty}
\frac{I^{\rm g}(\Omega)_{a'b'n_1n_2}
I^{\rm g'}(\Omega-\oee_{a'}+\oee_a)_{n_1n_2ab}
}
{(\oee_{a}+\oee_{b}-\oee_{n_1}-\oee_{n_2})
 (\Omega-\oee_{n_2}+\oee_{b'}+i0\oee_{n_2})}\,d\Omega \nonumber
\\
&&
+
\frac{i}{2\pi}\int\limits_{-\infty}^{\infty}
\frac{I^{\rm g}(\Omega)_{b'a'n_1n_2}
I^{\rm g'}(\Omega-\oee_{a}+\oee_{a'})_{n_1n_2ba}
}
{(\oee_{a}+\oee_{b}-\oee_{n_1}-\oee_{n_2})
 (\Omega-\oee_{n_2}+\oee_{a'}+i0\oee_{n_2})}\,d\Omega
\left.\vphantom{\frac{8}{8}}\right\}
\,,
\end{eqnarray}
\begin{eqnarray}
F_{a'b'ab}^{(2)(\rm box,red)}
\label{boxred}
&=&
-\frac{1}{2}
\mathop{{\sum}}_{\rm g g'}
\mathop{{\sum}''}_{n_1 n_2}
\left\{ \vphantom{\frac{8}{8}}\right.
\frac{i}{2\pi}\int\limits_{-\infty}^{\infty}
\frac{I^{\rm g}(\Omega)_{a'b'n_1n_2}
I^{\rm g'}(\Omega-\oee_{a'}+\oee_a)_{n_1n_2ab}
}
{(\Omega-\oee_{n_2}+\oee_{b'}+i0\oee_{n_2})^2}\,d\Omega
 \nonumber \\
&&
+
\frac{i}{2\pi}\int\limits_{-\infty}^{\infty}
\frac{I^{\rm g}(\Omega)_{b'a'n_1n_2}
I^{\rm g'}(\Omega-\oee_{a}+\oee_{a'})_{n_1n_2ba}
}
{(\Omega-\oee_{n_2}+\oee_{a'}+i0\oee_{n_2})^2}\,d\Omega
\left. \vphantom{\frac{8}{8}}\right\}
\,,
\end{eqnarray}
\begin{eqnarray}
F_{a'b'ab}^{(2)(\rm cross,irr)}
\label{crossirr}
&=&
\mathop{{\sum}}_{\rm g g'}
\mathop{{\sum}'}_{n_1 n_2}
\left\{ \vphantom{\frac{8}{8}}\right.
\frac{i}{2\pi}\int\limits_{-\infty}^{\infty}
\frac{I^{\rm g}(\Omega)_{b'n_2n_1a}
I^{\rm g'}(\Omega-\oee_{a'}+\oee_a)_{n_1a'bn_2}
}
{(\oee_{n_2}-\oee_{n_1}-\oee_{a}+\oee_{b'})
 (\Omega-\oee_{n_2}+\oee_{a}+i0\oee_{n_2})}\,d\Omega \nonumber
\\
&&
+
\frac{i}{2\pi}\int\limits_{-\infty}^{\infty}
\frac{I^{\rm g}(\Omega)_{n_1b'an_2}
I^{\rm g'}(\Omega-\oee_{a}+\oee_{a'})_{a'n_2n_1b}
}
{(\oee_{n_2}-\oee_{n_1}+\oee_{a}-\oee_{b'})
 (\Omega-\oee_{n_2}+\oee_{b'}+i0\oee_{n_2})}\,d\Omega
\left.\vphantom{\frac{8}{8}}\right\}
\,,
\end{eqnarray}
\begin{eqnarray}
F_{a'b'ab}^{(2)(\rm cross,red)}
\label{crossred}
&=&
\mathop{{\sum}}_{\rm g g'}
\mathop{{\sum}''}_{n_1 n_2}
\frac{i}{2\pi}\int\limits_{-\infty}^{\infty}
\frac{I^{\rm g}(\Omega)_{b'n_2n_1a}
I^{\rm g'}(\Omega-\oee_{a'}+\oee_a)_{n_1a'bn_2}
}
{(\Omega-\oee_{n_2}+\oee_{a}+i0\oee_{n_2})^2}\,d\Omega
\,.
\end{eqnarray}
The prime at the summation symbols indicates that the proper reference
state
members are ejected.
The double prime indicates that
only the reference state members are retained.
In order to avoid division by zero in
Eq. (\ref{crossirr}) in the case
$a=b'$
and
$n_1=n_2$
one has to take the limit
$\oee_{n_1}\to \oee_{n_2}$
in both terms on the right-hand side.
Thus the singularities cancel.
It should be stressed that Eq. (\ref{crossred}) coincides with
the result of this limiting process.
From
Eqs. (\ref{boxred}) and (\ref{crossred})
it follows automatically, that the corrections vanish for
${\rm g=g'=c}$.
The case 
${\rm g=g'=c}$
corresponds to the Coulomb-Coulomb correction, the case
${\rm g=g'=t}$
determines the Breit-Breit and the case
${\rm g=c, g'=t}$
or
${\rm g=t, g'=c}$
refers to the Coulomb-Breit
interaction.
\par
\begin{figure}
\begin{center}
\end{center}
\centerline{ \mbox{ \epsfxsize=0.15\textwidth \epsffile{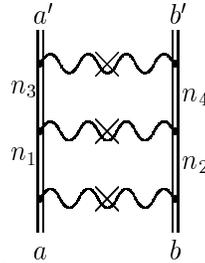} }}
\caption{
The third-order
``box'' Feynman graph.
The notations are the same as in Figs. 1, 2.
Here the wavy lines with the cross denote the
sum of the Coulomb and unretarded Breit
interaction.
}
\label{fig3}
\end{figure}

For high-$Z$ ions considered in this work
the third-order
contribution turns out to be small and it is sufficient 
to take into account its 
dominant part only.
Accordingly, we consider only the third-order Coulomb and
unretarded Breit ``box'' corrections.
The corresponding Feynman graph is displayed in
Fig.\,\ref{fig3}.
The formula for the irreducible part of
the third-order ``box'' correction can be
derived in the same manner as for the corrections given by 
Eqs. (\ref{ant}, \ref{boxirr}).
It takes the form
\begin{eqnarray}
F_{a'b'ab}^{\rm (3) (box,irr)}
\label{boxccc}
&=&
\mathop{{\sum}}_{\rm g g' g''}
\mathop{{\sum}'}_{n_1 n_2 n_3 n_4}
\frac{I_{a'b'n_3n_4}^{\rm g}I_{n_3n_4n_1n_2}^{\rm g'}
      I_{n_1n_2ab}^{\rm g''}}
     {(\oee_{n_3}+\oee_{n_4}-\oee_{a'}-\oee_{b'})
      (\oee_{n_1}+\oee_{n_2}-\oee_{a}-\oee_{b})}
\,,
\end{eqnarray}
where the prime indicates that the 
reference state contributions are excluded from
the summation.
Here the reference states are defined by the conditions 
$\oee_{n_1}+\oee_{n_2}=\oee_{a}+\oee_{b}$
or
$\oee_{n_3}+\oee_{n_4}=\oee_{a}+\oee_{b}$.
Applying the LPA to the graph in
Fig.\,\ref{fig3}
we derive the following expression for the reducible part
\begin{eqnarray}
F_{a'b'ab}^{ {(3)}{\rm (box,red)}}
\label{boxcccred}
&=&
\mathop{{\sum}}_{\rm g g' g''}
\mathop{{\sum}''}_{n_1 n_2 n_3 n_4}
{I_{a'b'n_3n_4}^{\rm g}I_{n_3n_4n_1n_2}^{\rm g'}
      I_{n_1n_2ab}^{\rm g''}}
 \nonumber \\
&&
\times
\left\{
  \frac{(-1)}
  {2(\oee_{n_3} + \oee_{n_4} - \oee_{a'} - \oee_{b'})^2}
+
  \frac{(-1)}
  {2(\oee_{n_1} + \oee_{n_2} - \oee_{a} - \oee_{b})^2}
\right\}
\,,
\end{eqnarray}
where the double prime indicates that the summation is running only
over the reference states.
The terms leading to vanishing denominators in
Eq. (\ref{boxcccred})
should be omitted.
\subsection{The three-electron configurations}
\label{theorythree}
Now we turn to three-electron ions.
Here we consider three-electron configurations with a closed 
$(1s)^2$ shell,
which can be
described by the wave function
\begin{eqnarray}
\Psi({\bm r}_{1}, {\bm r}_{2}, {\bm r}_{3})
\label{wavef3}
&=&
\frac{1}{\sqrt{3!}}
\sum\limits_{i,j,k=1,2,3}
\epsilon_{ijk}
\,
\psi_{i}({\bm r}_{1})
\psi_{j}({\bm r}_{2})
\psi_{k}({\bm r}_{3})
\,.
\end{eqnarray}
$\epsilon_{ijk}$
denotes the
Levi-Civita symbol
and
$\psi_1({\bm r})$,
$\psi_2({\bm r})$,
$\psi_3({\bm r})$
denote one-electron wave functions.
\par
As in the two-electron case we have to consider
corrections represented by the two-electron Feynman graphs
depicted in
Figs.\,\ref{fig1}-\ref{fig3}.
Their contribution to the energy shift is given by
\begin{eqnarray}
\Delta E(\{abc\})
\label{defabc1}
&=&
F_{ab;ab}+F_{bc;bc}+F_{ca;ca}
\,,
\end{eqnarray}
where
$F_{ab;cd}$
is given by
Eqs. (\ref{defmat}) and (\ref{ant}
-
\ref{boxcccred}).
The set
$\{abc\}$
is equal to the set
$\{1s_{+}, 1s_{-}, 2s_{1/2+}\}$
for the configuration
$(1s)^2 2s_{1/2}$
and to the set
$\{1s_{+}, 1s_{-}, 2p_{1/2+}\}$
for the configuration
$(1s)^2 2p_{1/2}$, respectively.
The symbol
$\pm$
refers to the different angular momentum projections.
\par
\begin{figure}
\begin{center}
\end{center}
\centerline{ \mbox{ \epsfxsize=0.2\textwidth \epsffile{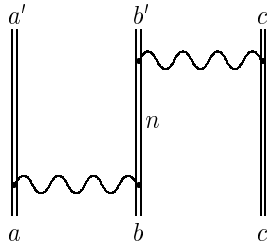} }}
\caption{
The second-order ``step'' graph for three-electron ions.
The notations are the same as in Figs. 1, 2.
}
\label{fig4}
\end{figure}
\begin{figure}
\begin{center}
\centerline{ \mbox{ \epsfxsize=0.5\textwidth \epsffile{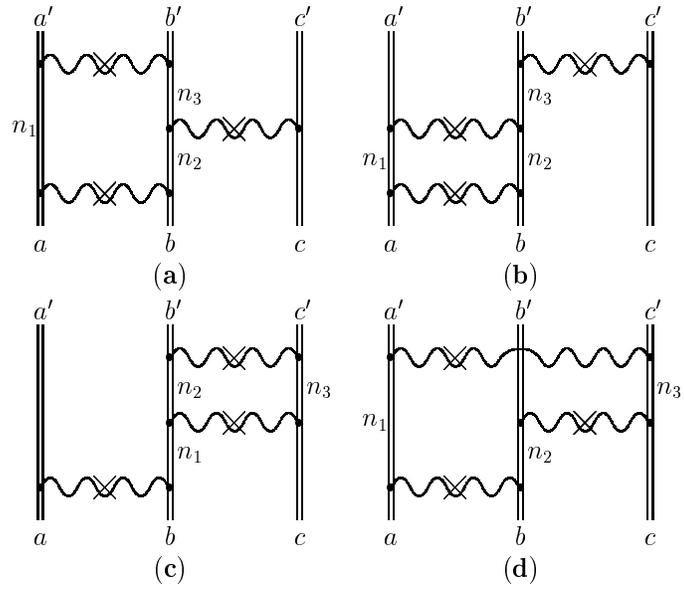} }}
\end{center}
\caption{
The third-order ``step-box'' graphs.
The wavy line with the cross denotes the sum of the Coulomb
and unretarded Breit interactions.
Otherwise, the notations are the same as in Figs. 1, 2, 3. 
}
\label{fig5}
\end{figure}

Besides the two-electron diagrams,
in three-electron problem we have to take into account
the additional
three-electron Feynman graphs depicted in
Figs.\,\ref{fig4} and \ref{fig5}.
The contribution of the three-electron graphs
can be calculated according to
\begin{eqnarray}
\Delta E{(\{abc\})}
\label{stepdet}
&=&
\sum_{{i', j', k' = 1, 2, 3}\atop{
  i, j, k = 1, 2, 3
  }}
\epsilon_{i' j' k'}
\,
\epsilon_{i j k}
F_{i' j' k' i j k}
\,,
\end{eqnarray}
where the indices
$1$, $2$, $3$
at
$F$
symbol must be replaced by
$a$, $b$, $c$
respectively,
i.e.
$F_{abcabc}\equiv F_{123123}$, etc.
Eq. (\ref{stepdet})
includes the contribution of the ``direct'' and all possible
``exchange'' diagrams which occur in the three-electron case.
\par
Expressions for 
$F_{a' b' c' a  b c}$
corresponding to the graph in
Fig.\,\ref{fig4}
are
\begin{eqnarray}
F_{a' b' c'
a  b  c }^{(2)({\rm step,irr})}
\label{stepirr}
&=&
\mathop{{\sum}}_{\rm g g'}
\mathop{{\sum}'}_{n}
\frac{I^{\rm g}(\oee_a - \oee_{a'})_{n a' b a}
  I^{\rm g'}(\oee_{c'} - \oee_c)_{b'c' n c}
  }{\oee_a + \oee_b - \oee_{a'} - \oee_{n}}
  \,,
\end{eqnarray}
\begin{eqnarray}
F_{a'b'c'abc}^{(2)(\rm step,red)}
\label{stepred}
&=&
\mathop{{\sum}}_{\rm g g'}
\mathop{{\sum}''}_{n}
\left.
\frac{\partial}{\partial\omega}
\left[
I^{\rm g}(\oee_{a}-\oee_{a'}+\omega)_{na'ba}
I^{\rm g'}(\oee_{c'}-\oee_{c}+\omega)_{b'c'nc}
\right]
\right|_{\textstyle\omega=0}
\,,
\end{eqnarray}
where the prime at the summation symbol  
indicates that the summation runs over all
$n$
except for the case when
the set of one-electron states
$\{a',n,c\}$
is equivalent to the set
$\{a,b,c\}$. The latter 
refers to reference states.
The double prime implies that the summation runs over the
reference states only.
As for the two-electron 
contributions we have
here
${\rm g, g' = c, t}$.
No reducible contribution arises for
${\rm g = g' = c}$.
\par
As it has been mentioned above, for the three-photon
corrections we take into account only their dominant parts,
i.e.,
the third-order Coulomb and
unretarded Breit ``box'' contributions.
The corresponding tree-electron Feynman graphs are
displayed in
Fig.\,\ref{fig5}.
The formulas for the irreducible and the reducible parts of
the third-order ``box'' correction are
derived in the same manner as in 
Eqs. (\ref{boxccc},\ref{boxcccred}).
The irreducible part can be expressed as
\begin{eqnarray}
F_{a' b' c' a  b c}^{(3)({\rm step-box,irr})}
\label{boxstepcccirr}
&=&
\mathop{{\sum}}_{\rm g g' g''}
\mathop{{\sum}'}_{n_1n_2n_3}
\frac{I^{\rm g}_{a'b'n_1n_3}I^{\rm g'}_{n_3c'n_2c}
  I^{\rm g''}_{n_1n_2a b}}
  {(\oee_{n_1} + \oee_{n_3} - \oee_{a'} - \oee_{b'})
   (\oee_{n_1} + \oee_{n_2} - \oee_{a} - \oee_{b})} \nonumber
\\
&&
+
2
\mathop{{\sum}}_{\rm g g' g''}
\mathop{{\sum}'}_{n_1n_2n_3}
\frac{I^{\rm g}_{b'c'n_3c}I^{\rm g'}_{a'n_3n_1n_2}
  I^{\rm g''}_{n_1n_2a b}}
  {(\oee_{n_1} + \oee_{n_2} - \oee_{a} - \oee_{b})
   (\oee_{n_3} + \oee_{a'} - \oee_{a} - \oee_{b})} \nonumber
\\
&&
+
\mathop{{\sum}}_{\rm g g' g''}
\mathop{{\sum}'}_{n_1n_2n_3}
\frac{I^{\rm g}_{a'c'n_1n_3}I^{\rm g'}_{b'n_3n_2c}
  I^{\rm g''}_{n_1n_2a b}}
  {(\oee_{n_1} + \oee_{n_2} - \oee_{a} - \oee_{b})
   (\oee_{n_1} + \oee_{n_3} - \oee_{a'} - \oee_{c'})}
\,,
\end{eqnarray}
where the prime at the summation symbols indicates that
the first summation does not run over states for which either
the set
$\{n_1,n_2,c\}$
or the set
$\{n_1,n_3,c'\}$
are equivalent to the set
$\{a,b,c\}$;
the second summation does not run over the states
for which the sets
$\{n_1,n_2,c\}$
or
$\{a',n_3,c\}$
are equivalent to the set
$\{a,b,c\}$
and
the third summation does not run over the states
for which the sets
$\{n_1,n_2,c\}$
or
$\{n_1,n_3,b'\}$
are equivalent to the set
$\{a,b,c\}$
(the cases of reference states).
The reducible part of the third-order ``step-box''  corrections
(see
Fig.\,\ref{fig5})
can be cast into the form
\begin{eqnarray}
F_{a' b' c' a\phantom{'} b\phantom{'}
c\phantom{'}}^{(3)({\rm step-box,red})}
&=&
\mathop{{\sum}}_{\rm g g' g''}
\mathop{{\sum}''}_{n_1n_2n_3}
{I^{\rm g}_{a'b'n_1n_3}I^{\rm g'}_{n_3c'n_2c}
  I^{\rm g''}_{n_1n_2a b}}
\nonumber \\
&&
\times
\left\{
  \frac{(-1)}
  {2(\oee_{n_1} + \oee_{n_3} - \oee_{a'} - \oee_{b'})^2}
+
  \frac{(-1)}
  {2(\oee_{n_1} + \oee_{n_2} - \oee_{a} - \oee_{b})^2}
\right\}
 \nonumber\\
&&
+
2
\mathop{{\sum}''}_{n_1n_2n_3}
{I^{\rm g}_{b'c'n_3c}I^{\rm g'}_{a'n_3n_1n_2}
  I^{\rm g''}_{n_1n_2a b}}
\nonumber \\
&&
\times
\left\{
  \frac{(-1)}
  {2(\oee_{n_1} + \oee_{n_2} - \oee_{a} - \oee_{b})^2}
+
  \frac{(-1)}
  {2(\oee_{n_3} + \oee_{a'} - \oee_{a} - \oee_{b})^2}
\right\}
 \nonumber\\
&&
+
\mathop{{\sum}''}_{n_1n_2n_3}
{I^{\rm g}_{a'c'n_1n_3}I^{\rm g'}_{b'n_3n_2c}
  I^{\rm g''}_{n_1n_2a b}}
 \nonumber\\
&&
\times
\left\{
  \frac{(-1)}
  {2(\oee_{n_1} + \oee_{n_2} - \oee_{a} - \oee_{b})^2}
+
  \frac{(-1)}
  {2(\oee_{n_1} + \oee_{n_3} - \oee_{a'} - \oee_{c'})^2}
\right\}
\,,
\end{eqnarray}
where the double prime at the summation symbols indicates that the
summations run over the corresponding reference states only 
(see the explanations for Eq. (\ref{boxstepcccirr})).
It becomes evident that the contributions
due to the graphs
Fig.\,\ref{fig5}b, c
are equal.
Therefore, we account for them by taking twice 
the contribution 
of the graph Fig.\,\ref{fig5}b.
\section{Numerical results and discussion}
\label{results}
The major result of the present work consists in the
calculation of the two- and three-photon
exchange corrections to the energy levels of two-electron
configurations
$2 {}^1\!S_{0}$,
$2 {}^3\!P_{0}$,
$2 {}^3\!S_{1}$
and three-electron configurations
$(1s)^2 2s_{1/2}$,
$(1s)^2 2p_{1/2}$.
The two-photon exchange correction represents the leading part of
the perturbation theory in second order.
Accordingly,
the main uncertainty of the theoretical values
calculated earlier 
has been due to this correction.
\par
In order to represent the Coulomb
potential of the nucleus we employ
a Fermi model for the nuclear density distribution
\begin{eqnarray}
\rho(r)
&=&
\frac{N}{1+\exp[(r-c)/a]}
\,,
\end{eqnarray}
where
$N$
is a normalization constant,
$a=0.5350\,\mbox{fm}$
and
$c$
is deduced via the equation
\begin{eqnarray}
4 \pi
\int_{0}^{\infty}
\rho(r) r^4
\,d r
&=&
\langle{r^2}\rangle
\,,
\end{eqnarray}
where
$\langle{r^2}\rangle^{1/2}$
is the root-mean-square nuclear radius.
In Table
\ref{tx8}
we also display the values for the nuclear root-mean-square radii
employed in this work. They have been 
taken from \cite{beier01}.
For nuclei with charge numbers
$Z$ not presented in
\cite{beier01} we utilize 
the empirical formula
\cite{johnson85}
\begin{eqnarray}
\langle{r^2}\rangle^{1/2}
&=&
(0.836\,A^{1/3}+0.570)\,\, \mbox{\rm fm}\, ,
\end{eqnarray}
where
$A$
is the atomic mass number.
\par
The results of our calculation of the two-photon exchange
correction are presented in Tables
\ref{tx1x1s0x1},
\ref{tx1x1s0x2},
\ref{tx1x3p0x1},
\ref{tx1x3p0x2},
\ref{tx1x3s1x1},
\ref{tx1x3s1x2}
for two-electron configurations
and in Tables
\ref{tx3xs12x1},
\ref{tx3xs12x2},
\ref{tx3xp12x1},
\ref{tx3xp12x2}
for three-electron configurations, respectively.
Our calculation is performed rigorously within the framework of QED.
For reasons of clarity
the corresponding corrections refering to 
contributions of the Feynman graphs in
Figs. \ref{fig2} and \ref{fig4} are also listed separately in these tables.
For details concerning the numerical procedure we refer to
\cite{andreev01}.
The accuracy of the present calculations is
on the level of about
$0.0001\, \mbox{\rm a.u}$.
\par
We also have taken into account the dominant part of
the three-photon exchange correction.
Details of the approximation made were given in
Sec. \ref{theory}.
The results of the calculation for the three-photon exchange
correction  are presented in Tables
\ref{tx2x1s0x1},
\ref{tx2x1s0x2},
\ref{tx2x3p0x1},
\ref{tx2x3p0x2},
\ref{tx2x3s1x1},
\ref{tx2x3s1x2}
for two-electron configurations and in Tables
\ref{tx4xs12x1},
\ref{tx4xs12x2},
\ref{tx4xp12x1},
\ref{tx4xp12x2}
for three-electron configurations, respectively.
Again, the contribution due to the exchange of the various
photons are compiled separately in the tables.
The correction caused by the exchange of three Breit photons is
not included since it was found to be less than
$0.001\, \mbox{\rm eV}$.
In view of the approximation used to evaluate the three-photon
exchange correction
these values are given within an inaccuracy of about
$10\, \mbox{\%}$ 
\cite{andreev01}.
\par
In Tables
\ref{tx5x1s0x1},
\ref{tx5x1s0x2},
\ref{tx5x1s0x3},
\ref{tx5x3p0x1},
\ref{tx5x3p0x2},
\ref{tx5x3p0x3},
\ref{tx5x3s1x1},
\ref{tx5x3s1x2},
\ref{tx5x3s1x3}
we collect all available corrections to the energy levels of
the two-electron configurations under consideration.
In order to compare our numerical data 
for the two-electron configurations
$2 {}^3\!P_{0}$
and
$2 {}^3\!S_{1}$ with other results in the literature  
we also provide values for the two-photon exchange correction
as it has been 
derived in
\cite{mohr00}.
We find that the data presented in
\cite{mohr00}
deviate from our results by not more than
$0.0003\, \mbox{a.u}$.
\par
Values for the energy of
the three-electron configurations are presented in 
Tables
\ref{tx6xs12x1},
\ref{tx6xs12x2},
\ref{tx6xs12x3},
\ref{tx6xp12x1},
\ref{tx6xp12x2},
\ref{tx6xp12x3}.
Comparing our results
for the two-photon exchange correction with data presented in
Ref. \cite{yerokhin01},
we achieved a very good agreement for the
$(1s)^2 2s_{1/2}$
configuration.
However, for the
$(1s)^2 2p_{1/2}$
level we find a discrepancy of about 
$0.0035\, \mbox{a.u.}$
for 
$Z=60, 70$.
The three-photon exchange correction is compared with
the results obtained in
\cite{zherebtsov00}.
In
\cite{zherebtsov00} the
exchange of two and three Breit photons has been neglected.
\par
The numbers for the recoil correction included in the tables 
for the total level energies of
two- and three-electron configurations 
are abtained by interpolation for those
$Z$
values 
not calculated in the refered paper.
The data for the self-energy (SE) screening and 
vacuum polarization (VP) screening of three-electron
configurations
as well as for the VP screening of two-electron configurations
have been obtained via a similar interpolation.
Results for the
SE screening corrections of
two-electron configurations
have been obtained according to a procedure which is 
based on the results provided in
\cite{indelicato01}. 
In particular we refer to Table II of 
Ref. \cite{indelicato01},
where the self-energy screening function
$f(Z\alpha)$
for K- and L-shell single-electron states have been presented.
From these values one can deduce the corresponding
self-energy shift of a single-electron state due to the screening 
effect of another single electron state.
E.g., we may denote by 
$E^{\rm{1s\, by\, 2s}}$
the screening correction to the 
$1s$-electron self-energy shift due to the 
$2s$-electron state and by
$E^{\rm{2s\, by\, 1s}}$
the screening correction to the
$2s$-self energy originating from the 
$1s$-electron state, respectively.
Accordingly, we suppose that the sum of the SE screening corrections
$E^{{}^1\!S_0 \,{\rm scr}}$
and
$E^{{}^3\!S_1 \,{\rm scr}}$
for the 
${}^1\!S_0$
and
${}^3\!S_1$
configuration is represented by
$E^{\rm{1s\, by\, 2s}}+E^{\rm{2s\, by\, 1s}} 
= 
E^{{}^1\!S_0 \,{\rm scr}}+E^{{}^3\!S_1 \,{\rm scr}}$.
Then we suppose that
$E^{{}^1\!S_0\, \rm{exch}}/E^{{}^3\!S_1\, \rm{exch}}
=
E^{{}^1\!S_0 \,{\rm scr}}/E^{{}^3\!S_1 \,{\rm scr}}$, where
$E^{{}^1\!S_0\, \rm{exch}}$
and
$E^{{}^3\!S_1\, \rm{exch}}$
are the first-order interelectron interaction corrections
for the corresponding configurations.
For the
${}^3\!P_0$ configuration we define
$E^{\rm{1s\, by\, 2p_{1/2}}}+E^{\rm{2p_{1/2}\, by\, 1s}}
=
E^{{}^3\!P_0 \,{\rm scr}}$.
\par
In Table
\ref{tx7x1}
we present the total values for the energy levels of
the two-electron configurations derived
in this paper and compare them with the results of
\cite{plante94}
and
\cite{drake88}, respectively.
The differences between the energy levels
are also listed in that table.
We should note that in 
Ref. \cite{drake88,plante94},
different approaches
have been employed,
i.e., the relativistic-all-order-theory (AO)
\cite{plante94}
and the unified theory (UT)
\cite{drake88}.
Compared with the rigorous QED approach, these theories involve
several approximations, i.e., neglection of 
(i) negative energy states,
(ii) crossed photon contributions, and
(iii) exact retardation effects.
However, they account for some part of higher-order interelectron
interaction corrections.
Accodingly, for highly-charged ions
the total data derived in Tables
\ref{tx5x1s0x1},
\ref{tx5x1s0x2},
\ref{tx5x3p0x1},
\ref{tx5x3p0x2},
\ref{tx5x3s1x1},
\ref{tx5x3s1x2},
\ref{tx6xs12x1},
\ref{tx6xs12x2},
\ref{tx6xp12x1},
\ref{tx6xp12x2}
provide the most accurate theoretical predictions for the energy levels
at  present.
\par
From the results presented in
Table
\ref{tx7x1}
one can conclude that the configurations
$2 {}^1\!S_{0}$
and
$2 {}^3\!P_{0}$
cross within the interval
$60<Z<70$.
Experimental investigation of PNC effects in heliumlike ions
requires a precise knowledge of the energy difference
between these levels at
$Z=63$
\cite{labzowsky01}.
UT theory
\cite{drake88}
predicts a value
for this difference of about
$0.168\, \mbox{eV}$, while 
the calculation presented in this paper gives a larger value of
$0.593\, \mbox{eV}$.
However, our calculations predict that the crossing of these levels
takes place near
$Z=66$
with an energy  difference of about 
$-0.016\, \mbox{eV}$.
Nevertheless, the He-like Eu ion
($Z=63$)
seems most suitable for the search of PNC effects \cite{labzowsky01}.
We also investigated the splitting
$E(2{}^1\!S_0)-E(2{}^3\!P_0)$
for two isotopes
${}^{151}_{\sr 63}$Eu
and
${}^{153}_{\sr 63}$Eu
and obtained an energy difference
$0.001 \mbox{eV}$,
which does not change the conclusions made in
\cite{labzowsky01}.
The present calculation also indicates that the other crossing point
can be expected to be close to
$Z=89,90$.
\section{Acknowledgements}
\label{acknowledgement}
The authors are indebted to Prof. W. Nagel from the computer center of 
the TU Dresden for providing access to the all 
necessary computer facilities.
O.A. is grateful to the TU Dresden for the
hospitality during his visits in 2001 and 2002 and to the DFG for 
financial support.
The work of O.A. and L.N. was supported by the RFBR Grant
No. 02-02-16578
and by Minobrazovanie grant No. E00-3.1-7.
G.P. and G.S. acknowledge financial support from BMBF, DFG and GSI.
%
%
%
%
%
%

%
%

\newpage

%
%
%
%
\begin{table}
\caption{The values of nuclear root-mean-square
   radii employed in this work.} 
\begin{tabular}{lrrrrrrrrrrrr} 
              &${}^{142}_{\sr 60}$Nd 
              &${}^{152}_{\sr 62}$Sm  
              &${}^{153}_{\sr 63}$Eu                
              &${}^{158}_{\sr 64}$Gd 
              &${}^{159}_{\sr 65}$Tb 
              &${}^{164}_{\sr 66}$Dy 
              &${}^{166}_{\sr 68}$Er 
              &${}^{174}_{\sr 70}$Yb 
              &${}^{202}_{\sr 80}$Hg 
              &${}^{231}_{\sr 91}$Pa 
              &${}^{238}_{\sr 92}$U 
              &${}^{237}_{\sr 93}$Np\\ 
 $\langle{r^2}\rangle^{1/2} \mbox{\rm (fm)}$&$4.914$&$5.032$&$5.041$&$5.090$&$5.099$&$5.224$&$5.165$&$5.317$&$5.467$&$5.700$&$5.860$&$5.744$\\ 
 \end{tabular} 
 \label{tx8} 
 \end{table} 


\begin{table} 
\caption{Different contributions to the second-order interelectron 
interaction for the two-electron configuration 
$1s_{1/2}2s_{1/2}{}^1\!S_0$ (eV). 
The numbers in the Table present 
the ionization energy of the 
$2s_{1/2}$ 
electron with the opposite sign.} 
\begin{tabular}{lrrrrrrrrrr} 
 Contribution&$Z=\;$60\phantom{88}&62\phantom{88} 
 &63\phantom{88}&64\phantom{88}&65\phantom{88} 
 &66\phantom{88}&68\phantom{88}&70\phantom{88}\\ 
 \hline 
 Coulomb-Coulomb&&&&\\ 
      \quad$\Delta E^{\mathrm{box,irr}}$&$    -3.929$&$    -3.996$&$    -4.032$&$    -4.069$&$    -4.107$&$    -4.146$&$    -4.230$&$    -4.314$\\ 
    \quad$\Delta E^{\mathrm{cross,irr}}$&$     0.019$&$     0.021$&$     0.022$&$     0.023$&$     0.024$&$     0.025$&$     0.028$&$     0.030$\\ 
        \quad$\Delta E^{\mathrm{total}}$&$    -3.909$&$    -3.975$&$    -4.010$&$    -4.046$&$    -4.083$&$    -4.120$&$    -4.203$&$    -4.284$\\ 
 Coulomb-Breit&&&&\\ 
      \quad$\Delta E^{\mathrm{box,irr}}$&$    -1.195$&$    -1.287$&$    -1.335$&$    -1.384$&$    -1.435$&$    -1.487$&$    -1.595$&$    -1.709$\\ 
      \quad$\Delta E^{\mathrm{box,red}}$&$     0.492$&$     0.534$&$     0.555$&$     0.577$&$     0.600$&$     0.624$&$     0.673$&$     0.726$\\
\end{tabular} 
 \label{tx1x1s0} 
 \label{tx1x1s0x1} 
 \end{table} 


\begin{table} 
\caption{Different contributions to the second-order interelectron 
interaction for the two-electron configuration 
$1s_{1/2}2s_{1/2}{}^1\!S_0$ (eV). 
The numbers in the Table present 
the ionization energy of the 
$2s_{1/2}$ 
electron with the opposite sign.} 
 \begin{tabular}{lrrrrrrrrrr} 
 Contribution&$Z=\;$80\phantom{88}&91\phantom{88} 
 &92\phantom{88}&93\phantom{88}\\ 
 \hline 
 Coulomb-Coulomb&&&&\\ 
      \quad$\Delta E^{\mathrm{box,irr}}$&$    -4.845$&$    -5.676$&$    -5.768$&$    -5.867$\\ 
    \quad$\Delta E^{\mathrm{cross,irr}}$&$     0.046$&$     0.071$&$     0.074$&$     0.077$\\ 
        \quad$\Delta E^{\mathrm{total}}$&$    -4.799$&$    -5.605$&$    -5.695$&$    -5.790$\\ 
 Coulomb-Breit&&&&\\ 
      \quad$\Delta E^{\mathrm{box,irr}}$&$    -2.375$&$    -3.391$&$    -3.499$&$    -3.618$\\ 
      \quad$\Delta E^{\mathrm{box,red}}$&$     1.037$&$     1.516$&$     1.569$&$     1.624$\\ 
    \quad$\Delta E^{\mathrm{cross,irr}}$&$    -0.108$&$    -0.138$&$    -0.140$&$    -0.145$\\ 
    \quad$\Delta E^{\mathrm{cross,red}}$&$    -0.146$&$    -0.232$&$    -0.242$&$    -0.252$\\ 
        \quad$\Delta E^{\mathrm{total}}$&$    -1.592$&$    -2.245$&$    -2.312$&$    -2.391$\\ 
 Breit-Breit&&&&\\ 
      \quad$\Delta E^{\mathrm{box,irr}}$&$    -0.302$&$    -0.508$&$    -0.528$&$    -0.554$\\ 
      \quad$\Delta E^{\mathrm{box,red}}$&$     0.127$&$     0.228$&$     0.239$&$     0.251$\\ 
    \quad$\Delta E^{\mathrm{cross,irr}}$&$     0.055$&$     0.087$&$     0.093$&$     0.098$\\ 
    \quad$\Delta E^{\mathrm{cross,red}}$&$     0.007$&$     0.013$&$     0.014$&$     0.014$\\ 
        \quad$\Delta E^{\mathrm{total}}$&$    -0.113$&$    -0.181$&$    -0.185$&$    -0.190$\\ 
 Total&&&&\\ 
                         \quad$\Delta E$&$    -6.504$&$    -8.032$&$    -8.184$&$    -8.371$\\ 
 \end{tabular} 
 \label{tx1x1s0x2} 
 \end{table} 


\begin{table} 
\caption{Different contributions to the second-order interelectron 
interaction for the two-electron configuration 
$1s_{1/2}2p_{1/2}{}^3\!P_0$ (eV). 
The numbers in the Table present 
the ionization energy of the 
$2p_{1/2}$ 
electron with the opposite sign.} 
 \begin{tabular}{lrrrrrrrrrr} 
 Contribution&$Z=\;$60\phantom{88}&62\phantom{88} 
 &63\phantom{88}&64\phantom{88}&65\phantom{88} 
 &66\phantom{88}&68\phantom{88}&70\phantom{88}\\ 
 \hline 
 Coulomb-Coulomb&&&&\\ 
      \quad$\Delta E^{\mathrm{box,irr}}$&$    -2.748$&$    -2.813$&$    -2.847$&$    -2.882$&$    -2.918$&$    -2.956$&$    -3.035$&$    -3.118$\\ 
    \quad$\Delta E^{\mathrm{cross,irr}}$&$     0.006$&$     0.007$&$     0.007$&$     0.007$&$     0.008$&$     0.009$&$     0.010$&$     0.012$\\ 
        \quad$\Delta E^{\mathrm{total}}$&$    -2.743$&$    -2.806$&$    -2.840$&$    -2.875$&$    -2.910$&$    -2.947$&$    -3.024$&$    -3.107$\\ 
 Coulomb-Breit&&&&\\ 
      \quad$\Delta E^{\mathrm{box,irr}}$&$    -1.162$&$    -1.259$&$    -1.310$&$    -1.363$&$    -1.417$&$    -1.473$&$    -1.587$&$    -1.708$\\ 
      \quad$\Delta E^{\mathrm{box,red}}$&$     0.048$&$     0.053$&$     0.056$&$     0.059$&$     0.062$&$     0.065$&$     0.071$&$     0.079$\\
\end{tabular} 
 \label{tx1x3p0} 
 \label{tx1x3p0x1} 
 \end{table} 


\begin{table} 
\caption{Different contributions to the second-order interelectron 
interaction for the two-electron configuration 
$1s_{1/2}2p_{1/2}{}^3\!P_0$ (eV). 
The numbers in the Table present 
the ionization energy of the 
$2p_{1/2}$ 
electron with the opposite sign.} 
 \begin{tabular}{lrrrrrrrrrr} 
 Contribution&$Z=\;$80\phantom{88}&91\phantom{88} 
 &92\phantom{88}&93\phantom{88}\\ 
 \hline 
 Coulomb-Coulomb&&&&\\ 
      \quad$\Delta E^{\mathrm{box,irr}}$&$    -3.636$&$    -4.472$&$    -4.567$&$    -4.667$\\ 
    \quad$\Delta E^{\mathrm{cross,irr}}$&$     0.022$&$     0.042$&$     0.044$&$     0.047$\\ 
        \quad$\Delta E^{\mathrm{total}}$&$    -3.614$&$    -4.430$&$    -4.523$&$    -4.620$\\ 
 Coulomb-Breit&&&&\\ 
      \quad$\Delta E^{\mathrm{box,irr}}$&$    -2.475$&$    -3.670$&$    -3.810$&$    -3.949$\\ 
      \quad$\Delta E^{\mathrm{box,red}}$&$     0.124$&$     0.199$&$     0.208$&$     0.216$\\ 
    \quad$\Delta E^{\mathrm{cross,irr}}$&$    -0.053$&$    -0.070$&$    -0.072$&$    -0.073$\\ 
    \quad$\Delta E^{\mathrm{cross,red}}$&$    -0.187$&$    -0.302$&$    -0.310$&$    -0.323$\\ 
        \quad$\Delta E^{\mathrm{total}}$&$    -2.590$&$    -3.842$&$    -3.985$&$    -4.128$\\ 
 Breit-Breit&&&&\\ 
      \quad$\Delta E^{\mathrm{box,irr}}$&$    -0.377$&$    -0.695$&$    -0.743$&$    -0.790$\\ 
      \quad$\Delta E^{\mathrm{box,red}}$&$     0.011$&$     0.023$&$     0.025$&$     0.026$\\ 
    \quad$\Delta E^{\mathrm{cross,irr}}$&$    -0.036$&$    -0.064$&$    -0.066$&$    -0.068$\\ 
    \quad$\Delta E^{\mathrm{cross,red}}$&$     0.008$&$     0.017$&$     0.018$&$     0.020$\\ 
        \quad$\Delta E^{\mathrm{total}}$&$    -0.394$&$    -0.718$&$    -0.766$&$    -0.811$\\ 
 Total&&&&\\ 
                         \quad$\Delta E$&$    -6.598$&$    -8.991$&$    -9.274$&$    -9.560$\\ 
 \end{tabular} 
 \label{tx1x3p0x2} 
 \end{table} 


\begin{table} 
\caption{Different contributions to the second-order interelectron 
interaction for the two-electron configuration 
$1s_{1/2}2s_{1/2}{}^3\!S_1$ (eV). 
The numbers in the Table present 
the ionization energy of the 
$2s_{1/2}$ 
electron with the opposite sign.} 
 \begin{tabular}{lrrrrrrrrrr} 
 Contribution&$Z=\;$60\phantom{88}&62\phantom{88} 
 &63\phantom{88}&64\phantom{88}&65\phantom{88} 
 &66\phantom{88}&68\phantom{88}&70\phantom{88}\\ 
 \hline 
 Coulomb-Coulomb&&&&\\ 
      \quad$\Delta E^{\mathrm{box,irr}}$&$    -1.535$&$    -1.555$&$    -1.565$&$    -1.575$&$    -1.586$&$    -1.597$&$    -1.620$&$    -1.643$\\ 
    \quad$\Delta E^{\mathrm{cross,irr}}$&$     0.001$&$     0.002$&$     0.002$&$     0.002$&$     0.002$&$     0.002$&$     0.002$&$     0.003$\\ 
        \quad$\Delta E^{\mathrm{total}}$&$    -1.534$&$    -1.553$&$    -1.563$&$    -1.574$&$    -1.584$&$    -1.595$&$    -1.617$&$    -1.641$\\ 
 Coulomb-Breit&&&&\\ 
      \quad$\Delta E^{\mathrm{box,irr}}$&$    -0.006$&$    -0.007$&$    -0.007$&$    -0.007$&$    -0.008$&$    -0.008$&$    -0.008$&$    -0.008$\\ 
      \quad$\Delta E^{\mathrm{box,red}}$&$    -0.002$&$    -0.003$&$    -0.003$&$    -0.003$&$    -0.003$&$    -0.003$&$    -0.003$&$    -0.004$\\
\end{tabular} 
 \label{tx1x3s1} 
 \label{tx1x3s1x1} 
 \end{table} 


\begin{table} 
\caption{Different contributions to the second-order interelectron 
interaction for the two-electron configuration 
$1s_{1/2}2s_{1/2}{}^3\!S_1$ (eV). 
The numbers in the Table present 
the ionization energy of the 
$2s_{1/2}$ 
electron with the opposite sign.} 
 \begin{tabular}{lrrrrrrrrrr} 
 Contribution&$Z=\;$80\phantom{88}&91\phantom{88} 
 &92\phantom{88}&93\phantom{88}\\ 
 \hline 
 Coulomb-Coulomb&&&&\\ 
      \quad$\Delta E^{\mathrm{box,irr}}$&$    -1.785$&$    -1.995$&$    -2.017$&$    -2.041$\\ 
    \quad$\Delta E^{\mathrm{cross,irr}}$&$     0.005$&$     0.008$&$     0.009$&$     0.009$\\ 
        \quad$\Delta E^{\mathrm{total}}$&$    -1.780$&$    -1.986$&$    -2.009$&$    -2.032$\\ 
 Coulomb-Breit&&&&\\ 
      \quad$\Delta E^{\mathrm{box,irr}}$&$    -0.010$&$    -0.011$&$    -0.012$&$    -0.012$\\ 
      \quad$\Delta E^{\mathrm{box,red}}$&$    -0.006$&$    -0.010$&$    -0.011$&$    -0.011$\\ 
    \quad$\Delta E^{\mathrm{cross,irr}}$&$    -0.009$&$    -0.015$&$    -0.015$&$    -0.015$\\ 
    \quad$\Delta E^{\mathrm{cross,red}}$&$     0.002$&$     0.004$&$     0.004$&$     0.004$\\ 
        \quad$\Delta E^{\mathrm{total}}$&$    -0.024$&$    -0.032$&$    -0.033$&$    -0.033$\\ 
 Breit-Breit&&&&\\ 
      \quad$\Delta E^{\mathrm{box,irr}}$&$     0.004$&$     0.005$&$     0.005$&$     0.006$\\ 
      \quad$\Delta E^{\mathrm{box,red}}$&$     0.000$&$     0.000$&$     0.000$&$     0.000$\\ 
    \quad$\Delta E^{\mathrm{cross,irr}}$&$     0.011$&$     0.017$&$     0.017$&$     0.017$\\ 
    \quad$\Delta E^{\mathrm{cross,red}}$&$     0.001$&$     0.002$&$     0.002$&$     0.002$\\ 
        \quad$\Delta E^{\mathrm{total}}$&$     0.015$&$     0.024$&$     0.024$&$     0.025$\\ 
 Total&&&&\\ 
                         \quad$\Delta E$&$    -1.789$&$    -1.995$&$    -2.018$&$    -2.040$\\ 
 \end{tabular} 
 \label{tx1x3s1x2} 
 \end{table} 

\begin{table}[t] 
\caption{Different contributions to the second-order interelectron 
interaction for the three-electron configuration 
$(1s)^2 2s_{1/2}$ (eV). 
The numbers in the Table present 
the ionization energy of the 
$2s_{1/2}$ 
electron with the opposite sign.} 
 \begin{tabular}{lrrrrrrrrrr} 
 Contribution&$Z=\;$60\phantom{88}&62\phantom{88} 
 &63\phantom{88}&64\phantom{88}&65\phantom{88} 
 &66\phantom{88}&68\phantom{88}&70\phantom{88}\\ 
 \hline 
 Coulomb-Coulomb&&&&\\ 
      \quad$\Delta E^{\mathrm{box,irr}}$&$    -4.267$&$    -4.330$&$    -4.363$&$    -4.397$&$    -4.432$&$    -4.468$&$    -4.546$&$    -4.622$\\ 
    \quad$\Delta E^{\mathrm{cross,irr}}$&$     0.012$&$     0.013$&$     0.013$&$     0.014$&$     0.015$&$     0.016$&$     0.017$&$     0.019$\\ 
     \quad$\Delta E^{\mathrm{step,irr}}$&$    -4.046$&$    -4.104$&$    -4.134$&$    -4.165$&$    -4.197$&$    -4.229$&$    -4.298$&$    -4.370$\\ 
        \quad$\Delta E^{\mathrm{total}}$&$    -8.302$&$    -8.421$&$    -8.484$&$    -8.548$&$    -8.614$&$    -8.682$&$    -8.826$&$    -8.974$\\ 
 Coulomb-Breit&&&&\\ 
      \quad$\Delta E^{\mathrm{box,irr}}$&$    -0.606$&$    -0.653$&$    -0.678$&$    -0.703$&$    -0.729$&$    -0.755$&$    -0.809$&$    -0.866$\\
\end{tabular} 
 \label{tx3xs12} 
 \label{tx3xs12x1} 
 \end{table}


\begin{table}[t] 
\caption{Different contributions to the second-order interelectron 
interaction for the three-electron configuration 
$(1s)^2 2s_{1/2}$ (eV). 
The numbers in the Table present 
the ionization energy of the 
$2s_{1/2}$ 
electron with the opposite sign.} 
 \begin{tabular}{lrrrrrrrrrr} 
 Contribution&$Z=\;$80\phantom{88}&91\phantom{88} 
 &92\phantom{88}&93\phantom{88}\\ 
 \hline 
 Coulomb-Coulomb&&&&\\ 
      \quad$\Delta E^{\mathrm{box,irr}}$&$    -5.100$&$    -5.830$&$    -5.910$&$    -5.995$\\ 
    \quad$\Delta E^{\mathrm{cross,irr}}$&$     0.030$&$     0.048$&$     0.050$&$     0.052$\\ 
     \quad$\Delta E^{\mathrm{step,irr}}$&$    -4.802$&$    -5.455$&$    -5.526$&$    -5.601$\\ 
        \quad$\Delta E^{\mathrm{total}}$&$    -9.872$&$   -11.238$&$   -11.386$&$   -11.543$\\ 
 Coulomb-Breit&&&&\\ 
      \quad$\Delta E^{\mathrm{box,irr}}$&$    -1.202$&$    -1.712$&$    -1.771$&$    -1.825$\\
\end{tabular} 
 \label{tx3xs12x2} 
 \end{table}


\begin{table}
\caption{Different contributions to the second-order interelectron 
interaction for the three-electron configuration 
$(1s)^2 2p_{1/2}$ (eV). 
The numbers in the Table present 
the ionization energy of the 
$2p_{1/2}$ 
electron with the opposite sign.} 
 \begin{tabular}{lrrrrrrrrrr} 
 Contribution&$Z=\;$60\phantom{88}&62\phantom{88} 
 &63\phantom{88}&64\phantom{88}&65\phantom{88} 
 &66\phantom{88}&68\phantom{88}&70\phantom{88}\\ 
 \hline 
 Coulomb-Coulomb&&&&\\ 
      \quad$\Delta E^{\mathrm{box,irr}}$&$    -8.253$&$    -8.276$&$    -8.294$&$    -8.317$&$    -8.345$&$    -8.378$&$    -8.456$&$    -8.552$\\ 
    \quad$\Delta E^{\mathrm{cross,irr}}$&$     0.013$&$     0.015$&$     0.016$&$     0.017$&$     0.018$&$     0.019$&$     0.021$&$     0.024$\\ 
     \quad$\Delta E^{\mathrm{step,irr}}$&$    -5.202$&$    -5.462$&$    -5.591$&$    -5.721$&$    -5.851$&$    -5.981$&$    -6.244$&$    -6.512$\\ 
        \quad$\Delta E^{\mathrm{total}}$&$   -13.443$&$   -13.723$&$   -13.870$&$   -14.022$&$   -14.178$&$   -14.340$&$   -14.679$&$   -15.040$\\ 
 Coulomb-Breit&&&&\\ 
      \quad$\Delta E^{\mathrm{box,irr}}$&$     0.226$&$     0.170$&$     0.137$&$     0.102$&$     0.066$&$     0.029$&$    -0.047$&$    -0.131$\\
\end{tabular} 
 \label{tx3xp12} 
 \label{tx3xp12x1} 
 \end{table} 


\begin{table}
\caption{Different contributions to the second-order interelectron 
interaction for the three-electron configuration 
$(1s)^2 2p_{1/2}$ (eV). 
The numbers in the Table present 
the ionization energy of the 
$2p_{1/2}$ 
electron with the opposite sign.} 
 \begin{tabular}{lrrrrrrrrrr} 
 Contribution&$Z=\;$80\phantom{88}&91\phantom{88} 
 &92\phantom{88}&93\phantom{88}\\ 
 \hline 
 Coulomb-Coulomb&&&&\\ 
      \quad$\Delta E^{\mathrm{box,irr}}$&$    -9.313$&$   -10.792$&$   -10.968$&$   -11.155$\\ 
    \quad$\Delta E^{\mathrm{cross,irr}}$&$     0.043$&$     0.078$&$     0.082$&$     0.087$\\ 
     \quad$\Delta E^{\mathrm{step,irr}}$&$    -7.984$&$   -10.080$&$   -10.309$&$   -10.548$\\ 
        \quad$\Delta E^{\mathrm{total}}$&$   -17.254$&$   -20.794$&$   -21.195$&$   -21.616$\\ 
 Coulomb-Breit&&&&\\ 
      \quad$\Delta E^{\mathrm{box,irr}}$&$    -0.619$&$    -1.366$&$    -1.446$&$    -1.532$\\
\end{tabular} 
 \label{tx3xp12x2} 
 \end{table}


\begin{table} 
\caption{Different contributions to the third-order interelectron 
interaction for the two-electron configuration 
$1s_{1/2}2s_{1/2}{}^1\!S_0$ (eV). 
The numbers in the Table present 
the ionization energy of the 
$2s_{1/2}$ 
electron with the opposite sign.} 
 \begin{tabular}{lrrrrrrrrrr} 
 Contribution&$Z=\;$60\phantom{88}&62\phantom{88} 
 &63\phantom{88}&64\phantom{88}&65\phantom{88} 
 &66\phantom{88}&68\phantom{88}&70\phantom{88}\\ 
 \hline 
                 Coulomb-Coulomb-Coulomb&$      0.010$&$      0.010$&$      0.010$&$      0.010$&$      0.010$&$      0.010$&$      0.010$&$      .010$\\ 
                   Coulomb-Coulomb-Breit&$      0.011$&$      0.011$&$      0.011$&$      0.012$&$      0.012$&$      0.012$&$      0.013$&$      .013$\\ 
                     Coulomb-Breit-Breit&$      0.005$&$      0.005$&$      0.005$&$      0.005$&$      0.005$&$      0.006$&$      0.006$&$      .006$\\ 
                   Total                &$      0.026$&$      0.026$&$      0.026$&$      0.027$&$      0.027$&$      0.028$&$      0.029$&$      .029$\\ 
 \end{tabular} 
 \label{tx2x1s0} 
 \label{tx2x1s0x1} 
 \end{table}


\begin{table} 
\caption{Different contributions to the third-order interelectron 
interaction for the two-electron configuration 
$1s_{1/2}2s_{1/2}{}^1\!S_0$ (eV). 
The numbers in the Table present 
the ionization energy of the 
$2s_{1/2}$ 
electron with the opposite sign.} 
 \begin{tabular}{lrrrrrrrrrr} 
 Contribution&$Z=\;$80\phantom{88}&91\phantom{88} 
 &92\phantom{88}&93\phantom{88}\\ 
 \hline 
                 Coulomb-Coulomb-Coulomb&$     0.012$&$     0.016$&$     0.016$&$     0.017$\\ 
                   Coulomb-Coulomb-Breit&$     0.016$&$     0.022$&$     0.022$&$     0.023$\\ 
                     Coulomb-Breit-Breit&$     0.008$&$     0.011$&$     0.012$&$     0.012$\\ 
                   Total                &$     0.036$&$     0.049$&$     0.050$&$     0.052$\\ 
 \end{tabular} 
 \label{tx2x1s0x2} 
 \end{table}


\begin{table} 
\caption{Different contributions to the third-order interelectron 
interaction for the two-electron configuration 
$1s_{1/2}2p_{1/2}{}^3\!P_0$ (eV). 
The numbers in the Table present 
the ionization energy of the 
$2p_{1/2}$ 
electron with the opposite sign.} 
 \begin{tabular}{lrrrrrrrrrr} 
 Contribution&$Z=\;$60\phantom{88}&62\phantom{88} 
 &63\phantom{88}&64\phantom{88}&65\phantom{88} 
 &66\phantom{88}&68\phantom{88}&70\phantom{88}\\ 
 \hline 
                 Coulomb-Coulomb-Coulomb&$    -0.005$&$     -0.004$&$     -0.004$&$     -0.004$&$     -0.004$&$     -0.003$&$     -0.003$&$     -0.003$\\ 
                   Coulomb-Coulomb-Breit&$     0.011$&$      0.012$&$      0.012$&$      0.013$&$      0.014$&$      0.014$&$      0.015$&$      0.015$\\ 
                     Coulomb-Breit-Breit&$     0.006$&$      0.005$&$      0.005$&$      0.006$&$      0.006$&$      0.006$&$      0.007$&$      0.008$\\ 
                   Total                &$     0.013$&$      0.013$&$      0.014$&$      0.015$&$      0.016$&$      0.017$&$      0.019$&$      0.020$\\ 
 \end{tabular} 
 \label{tx2x3p0} 
 \label{tx2x3p0x1} 
 \end{table} 


\begin{table} 
\caption{Different contributions to the third-order interelectron 
interaction for the two-electron configuration 
$1s_{1/2}2p_{1/2}{}^3\!P_0$ (eV). 
The numbers in the Table present 
the ionization energy of the 
$2p_{1/2}$ 
electron with the opposite sign.} 
 \begin{tabular}{lrrrrrrrrrr} 
 Contribution&$Z=\;$80\phantom{88}&91\phantom{88} 
 &92\phantom{88}&93\phantom{88}\\ 
 \hline 
                 Coulomb-Coulomb-Coulomb&$      0.000$&$      0.004$&$      0.004$&$      0.005$\\ 
                   Coulomb-Coulomb-Breit&$      0.021$&$      0.032$&$      0.033$&$      0.034$\\ 
                     Coulomb-Breit-Breit&$      0.013$&$      0.020$&$      0.021$&$      0.022$\\ 
                   Total                &$      0.034$&$      0.056$&$      0.058$&$      0.061$\\ 
 \end{tabular} 
 \label{tx2x3p0x2} 
 \end{table}


\begin{table} 
\caption{Different contributions to the third-order interelectron 
interaction for the two-electron configuration 
$1s_{1/2}2s_{1/2}{}^3\!S_1$ (eV). 
The numbers in the Table present 
the ionization energy of the 
$2s_{1/2}$ 
electron with the opposite sign.} 
 \begin{tabular}{lrrrrrrrrrr} 
 Contribution&$Z=\;$60\phantom{88}&62\phantom{88} 
 &63\phantom{88}&64\phantom{88}&65\phantom{88} 
 &66\phantom{88}&68\phantom{88}&70\phantom{88}\\ 
 \hline 
                 Coulomb-Coulomb-Coulomb&$     -0.001$&$     -0.001$&$     -0.001$&$     -0.001$&$     -0.001$&$     -0.001$&$     -0.001$&$     -0.001$\\ 
                   Coulomb-Coulomb-Breit&$      0.001$&$      0.001$&$      0.001$&$      0.001$&$      0.001$&$      0.001$&$      0.001$&$      0.001$\\ 
                     Coulomb-Breit-Breit&$      0.000$&$      0.000$&$      0.000$&$      0.000$&$      0.000$&$      0.000$&$      0.000$&$      0.000$\\ 
                   Total                &$     -0.001$&$     -0.001$&$     -0.001$&$     -0.001$&$      0.000$&$      0.000$&$      0.000$&$      0.000$\\ 
 \end{tabular} 
 \label{tx2x3s1} 
 \label{tx2x3s1x1} 
 \end{table}


\begin{table} 
\caption{Different contributions to the third-order interelectron 
interaction for the two-electron configuration 
$1s_{1/2}2s_{1/2}{}^3\!S_1$ (eV). 
The numbers in the Table present 
the ionization energy of the 
$2s_{1/2}$ 
electron with the opposite sign.} 
 \begin{tabular}{lrrrrrrrrrr} 
 Contribution&$Z=\;$80\phantom{88}&91\phantom{88} 
 &92\phantom{88}&93\phantom{88}\\ 
 \hline 
                 Coulomb-Coulomb-Coulomb&$     -0.001$&$      0.000$&$      0.000$&$      0.000$\\ 
                   Coulomb-Coulomb-Breit&$      0.001$&$      0.001$&$      0.001$&$      0.001$\\ 
                     Coulomb-Breit-Breit&$      0.000$&$      0.000$&$      0.000$&$      0.000$\\ 
                   Total                &$      0.000$&$      0.001$&$      0.001$&$      0.001$\\ 
 \end{tabular} 
 \label{tx2x3s1x2} 
 \end{table} 


\begin{table}[t] 
\caption{Different contributions to the third-order interelectron 
interaction for the three-electron configuration 
$(1s)^2 2s_{1/2}$ (eV). 
The numbers in the Table present 
the ionization energy of the 
$2s_{1/2}$ 
electron with the opposite sign.} 
 \begin{tabular}{lrrrrrrrrrr} 
 Contribution&$Z=\;$60\phantom{88}&62\phantom{88} 
 &63\phantom{88}&64\phantom{88}&65\phantom{88} 
 &66\phantom{88}&68\phantom{88}&70\phantom{88}\\ 
 \hline 
                 Coulomb-Coulomb-Coulomb&$      0.003$&$      0.004$&$      0.004$&$      0.005$&$      0.006$&$      0.006$&$      0.007$&$      0.008$\\ 
                   Coulomb-Coulomb-Breit&$      0.014$&$      0.014$&$      0.015$&$      0.016$&$      0.016$&$      0.016$&$      0.016$&$      0.017$\\ 
                     Coulomb-Breit-Breit&$      0.002$&$      0.003$&$      0.004$&$      0.004$&$      0.005$&$      0.006$&$      0.006$&$      0.006$\\ 
                        Total           &$      0.018$&$      0.021$&$      0.023$&$      0.025$&$      0.027$&$      0.028$&$      0.029$&$      0.031$\\ 
 \end{tabular} 
 \label{tx4xs12} 
 \label{tx4xs12x1} 
 \end{table}


\begin{table}[t] 
\caption{Different contributions to the third-order interelectron 
interaction for the three-electron configuration 
$(1s)^2 2s_{1/2}$ (eV). 
The numbers in the Table present 
the ionization energy of the 
$2s_{1/2}$ 
electron with the opposite sign.} 
 \begin{tabular}{lrrrrrrrrrr} 
 Contribution&$Z=\;$80\phantom{88}&91\phantom{88} 
 &92\phantom{88}&93\phantom{88}\\ 
 \hline 
                 Coulomb-Coulomb-Coulomb&$      0.015$&$      0.025$&$      0.026$&$      0.027$\\ 
                   Coulomb-Coulomb-Breit&$      0.021$&$      0.028$&$      0.029$&$      0.030$\\ 
                     Coulomb-Breit-Breit&$      0.007$&$      0.011$&$      0.011$&$      0.012$\\ 
                        Total           &$      0.043$&$      0.064$&$      0.066$&$      0.069$\\ 
 \end{tabular} 
 \label{tx4xs12x2} 
 \end{table}


\begin{table}[t] 
\caption{Different contributions to the third-order interelectron 
interaction for the three-electron configuration 
$(1s)^2 2p_{1/2}$ (eV). 
The numbers in the Table present 
the ionization energy of the 
$2p_{1/2}$ 
electron with the opposite sign.} 
 \begin{tabular}{lrrrrrrrrrr} 
 Contribution&$Z=\;$60\phantom{88}&62\phantom{88} 
 &63\phantom{88}&64\phantom{88}&65\phantom{88} 
 &66\phantom{88}&68\phantom{88}&70\phantom{88}\\ 
 \hline 
                 Coulomb-Coulomb-Coulomb&$      0.002$&$      0.005$&$      0.007$&$      0.009$&$      0.011$&$      0.013$&$      0.016$&$      0.019$\\ 
                   Coulomb-Coulomb-Breit&$      0.049$&$      0.051$&$      0.053$&$      0.054$&$      0.055$&$      0.056$&$      0.058$&$      0.059$\\ 
                     Coulomb-Breit-Breit&$      0.010$&$      0.011$&$      0.011$&$      0.011$&$      0.011$&$      0.012$&$      0.012$&$      0.012$\\ 
                        Total           &$      0.061$&$      0.067$&$      0.071$&$      0.074$&$      0.077$&$      0.080$&$      0.085$&$      0.090$\\ 
 \end{tabular} 
 \label{tx4xp12} 
 \label{tx4xp12x1} 
 \end{table} 


\begin{table}[t] 
\caption{Different contributions to the third-order interelectron 
interaction for the three-electron configuration 
$(1s)^2 2p_{1/2}$ (eV). 
The numbers in the Table present 
the ionization energy of the 
$2p_{1/2}$ 
electron with the opposite sign.} 
 \begin{tabular}{lrrrrrrrrrr} 
 Contribution&$Z=\;$80\phantom{88}&91\phantom{88} 
 &92\phantom{88}&93\phantom{88}\\ 
 \hline 
                 Coulomb-Coulomb-Coulomb&$      0.041$&$      0.079$&$      0.083$&$      0.088$\\ 
                   Coulomb-Coulomb-Breit&$      0.084$&$      0.116$&$      0.119$&$      0.124$\\ 
                     Coulomb-Breit-Breit&$      0.017$&$      0.030$&$      0.031$&$      0.033$\\ 
                        Total           &$      0.142$&$      0.225$&$      0.233$&$      0.245$\\ 
 \end{tabular} 
 \label{tx4xp12x2} 
 \end{table} 


\begin{table}
\caption{Different contributions to the total energy of the 
two-electron configuration 
$1s_{1/2}2s_{1/2} \,{}^1\! S_0$ (eV). 
The numbers in the Table present 
the ionization energy of the 
$2s_{1/2}$ 
electron with the opposite sign.} 
 \begin{tabular}{llrrrrc} 
 \multicolumn{2}{l}{Contribution}&$Z=\;$         60\phantom{88}& 
          62\phantom{88} 
 &         63\phantom{88}&         64\phantom{88} 
 &Ref.\\ 
 \hline 
          \multicolumn{2}{l}{Zero-order}&$-13063.004$&$-14015.106$&$-14506.589$&$-15008.567$&                            TW\\ 
   \multicolumn{2}{l}{Nuclear size (NS)}&$     0.928$&$     1.201$&$     1.342$&$     1.519$&                            TW\\ 
 \multicolumn{2}{l}{First-order }       &&&&&\\ 
       \multicolumn{2}{l}{interelectron}&$   438.812$&$   458.300$&$   468.260$&$   478.371$&                            TW\\ 
 \multicolumn{2}{l}{interaction}        &&&&&\\ 
 \multicolumn{2}{l}{Second-order}       &&&&&\\ 
       \multicolumn{2}{l}{interelectron}&$    -4.781$&$    -4.913$&$    -4.980$&$    -5.049$&                            TW\\
\end{tabular} 
 \label{tx5x1s0x1} 
 \end{table}


\begin{table}
\caption{Different contributions to the total energy of the 
two-electron configuration 
$1s_{1/2}2s_{1/2} \,{}^1\! S_0$ (eV). 
The numbers in the Table present 
the ionization energy of the 
$2s_{1/2}$ 
electron with the opposite sign.} 
 \begin{tabular}{llrrrrc} 
 \multicolumn{2}{l}{Contribution}&$Z=\;$         65\phantom{88}& 
          66\phantom{88} 
 &         68\phantom{88}&         70\phantom{88} 
 &Ref.\\ 
 \hline 
          \multicolumn{2}{l}{Zero-order}&$-15521.201$&$-16044.661$&$-17124.764$&$-18250.361$&                            TW\\ 
   \multicolumn{2}{l}{Nuclear size (NS)}&$     1.696$&$     1.968$&$     2.386$&$     3.099$&                            TW\\ 
 \multicolumn{2}{l}{First-order }       &&&&&\\ 
       \multicolumn{2}{l}{interelectron}&$   488.637$&$   499.061$&$   520.422$&$   542.484$&                            TW\\ 
 \multicolumn{2}{l}{interaction}        &&&&&\\ 
 \multicolumn{2}{l}{Second-order}       &&&&&\\ 
       \multicolumn{2}{l}{interelectron}&$    -5.120$&$    -5.194$&$    -5.350$&$    -5.515$&                            TW\\ 
         \multicolumn{2}{l}{interaction}&$          $&$          $&$          $&$          $&                              \\ 
 \multicolumn{2}{l}{Third-order}        &&&&&\\ 
       \multicolumn{2}{l}{interelectron}&$     0.027$&$     0.028$&$     0.029$&$     0.029$&                            TW\\ 
 \multicolumn{2}{l}{interaction}        &&&&&\\ 
                            SE with NS &&$    15.427$&$    16.358$&$    18.363$&$    20.589$&          \cite{mohr92,mohr93}\\ 
                            VP with NS &&$    -2.374$&$    -2.557$&$    -2.962$&$    -3.419$&       \cite{johnson85,soff88}\\ 
                          SE screening &&$    -0.335$&$    -0.351$&$    -0.382$&$    -0.413$&          \cite{indelicato01}\\ 
                          VP screening &&$     0.133$&$     0.139$&$     0.164$&$     0.188$&       \cite{beier01artemyev}\\ 
                              Recoil   &&$     0.060$&$     0.061$&$     0.064$&$     0.067$&             \cite{artemyev95}\\ 
               \multicolumn{2}{l}{Total}&$-15023.050$&$-15535.148$&$-16592.030$&$-17693.252$&                            TW\\ 
 \end{tabular} 
 \label{tx5x1s0x2} 
 \end{table}


\begin{table}
\caption{Different contributions to the total energy of the 
two-electron configuration 
$1s_{1/2}2s_{1/2} \,{}^1\! S_0$ (eV). 
The numbers in the Table present 
the ionization energy of the 
$2s_{1/2}$ 
electron with the opposite sign.} 
 \begin{tabular}{llrrrrc} 
 \multicolumn{2}{l}{Contribution}&$Z=\;$         80\phantom{88}& 
          91\phantom{88} 
 &         92\phantom{88}&         93\phantom{88} 
 &Ref.\\ 
 \hline 
          \multicolumn{2}{l}{Zero-order}&$ -24622.160$&$-33320.132$&$-34215.481$&$-35130.460$&                            TW\\ 
   \multicolumn{2}{l}{Nuclear size (NS)}&$     9.34$  &$      32.50$&$     37.76$ &$     40.94 $&                            TW\\ 
 \multicolumn{2}{l}{First-order }       &&&&&\\ 
       \multicolumn{2}{l}{interelectron}&$   665.38$  &$    832.758$&$   850.116$ &$    868.029$&                            TW\\ 
 \multicolumn{2}{l}{interaction}        &&&&&\\ 
 \multicolumn{2}{l}{Second-order}       &&&&&\\ 
       \multicolumn{2}{l}{interelectron}&$    -6.504$ &$    -8.032$&$    -8.184$ &$    -8.371$&                            TW\\ 
         \multicolumn{2}{l}{interaction}&$          $ &$          $&$   -8.2131$ &$          $& \cite{asen02}\\ 
 \multicolumn{2}{l}{Third-order}        &&&&&\\ 
       \multicolumn{2}{l}{interelectron}&$     0.038$ &$     0.049$&$     0.052$ &$     0.052$&                            TW\\ 
 \multicolumn{2}{l}{interaction}        &&&&&\\ 
                            SE with NS &&$   35.3911$ &$    62.226$&$    65.4183$&$    68.739$&          \cite{mohr92,mohr93}\\ 
                            VP with NS &&$    -6.900$ &$    -14.632$&$    -15.658$&$   -16.770$&       \cite{johnson85,soff88}\\ 
                          SE screening &&$    -0.651$ &$     -1.087$&$    -1.127$ &$    -1.167$&          \cite{indelicato01}\\ 
                          VP screening &&$     0.354$ &$      0.729$&$     0.777$&$    0.825$&       \cite{beier01artemyev}\\ 
                              Recoil   &&$    0.086 $ &$    0.122 $&$     0.127 $&$    0.132 $&             \cite{artemyev95}\\ 
               \multicolumn{2}{l}{Total}&$ -23925.626$&$-32415.499$&$-33286.200$&$-34178.051$&                            TW\\ 
 \end{tabular} 
 \label{tx5x1s0x3} 
 \end{table} 


\begin{table}
\caption{Different contributions to the total energy of the 
two-electron configuration 
$1s_{1/2}2p_{1/2} \,{}^3\! P_0$ (eV). 
The numbers in the Table present 
the ionization energy of the 
$2p_{1/2}$ 
electron with the opposite sign.} 
 \begin{tabular}{llrrrrc} 
 \multicolumn{2}{l}{Contribution}&$Z=\;$         60\phantom{88}& 
          62\phantom{88} 
 &         63\phantom{88}&         64\phantom{88} 
 &Ref.\\ 
 \hline 
          \multicolumn{2}{l}{Zero-order}&$-13063.004$&$-14015.106$&$-14506.589$&$-15008.567$&                            TW\\ 
   \multicolumn{2}{l}{Nuclear size (NS)}&$     0.038$&$     0.052$&$     0.061$&$     0.071$&                            TW\\ 
 \multicolumn{2}{l}{First-order }       &&&&&\\ 
       \multicolumn{2}{l}{interelectron}&$   447.329$&$   468.642$&$   479.586$&$   490.731$&                            TW\\ 
 \multicolumn{2}{l}{interaction}        &&&&&\\ 
 \multicolumn{2}{l}{Second-order}       &&&&&\\ 
       \multicolumn{2}{l}{interelectron}&$    -4.068$&$    -4.256$&$    -4.354$&$    -4.456$&                            TW\\ 
         \multicolumn{2}{l}{interaction}&$   -4.0645$&$          $&$          $&$          $&                 \cite{mohr00}\\ 
 \multicolumn{2}{l}{Third-order}        &&&&&\\ 
       \multicolumn{2}{l}{interelectron}&$     0.013$&$     0.013$&$     0.014$&$     0.015$&                            TW\\ 
 \multicolumn{2}{l}{interaction}        &&&&&\\ 
                            SE with NS &&$     0.298$&$     0.406$&$     0.469$&$     0.539$&          \cite{mohr92,mohr93}\\ 
                            VP with NS &&$    -0.098$&$    -0.124$&$    -0.139$&$    -0.155$&       \cite{johnson85,soff88}\\ 
                          SE screening &&$    -0.112$&$    -0.133$&$    -0.144$&$    -0.155$&          \cite{indelicato01}\\ 
                          VP screening &&$     0.035$&$     0.039$&$     0.042$&$     0.044$&       \cite{beier01artemyev}\\ 
                              Recoil   &&$     0.050$&$     0.051$&$     0.052$&$     0.052$&             \cite{artemyev95}\\ 
               \multicolumn{2}{l}{Total}&$-12619.519$&$-13550.416$&$-14031.002$&$-14521.881$&                            TW\\ 
 \end{tabular} 
 \label{tx5x3p0x1} 
 \end{table} 


\begin{table}
\caption{Different contributions to the total energy of the 
two-electron configuration 
$1s_{1/2}2p_{1/2} \,{}^3\! P_0$ (eV). 
The numbers in the Table present 
the ionization energy of the 
$2p_{1/2}$ 
electron with the opposite sign.} 
 \begin{tabular}{llrrrrc} 
 \multicolumn{2}{l}{Contribution}&$Z=\;$         65\phantom{88}& 
          66\phantom{88} 
 &         68\phantom{88}&         70\phantom{88} 
 &Ref.\\ 
 \hline 
          \multicolumn{2}{l}{Zero-order}&$-15521.201$&$-16044.661$&$-17124.764$&$-18250.361$&                            TW\\ 
   \multicolumn{2}{l}{Nuclear size (NS)}&$     0.083$&$     0.099$&$     0.129$&$     0.180$&                            TW\\ 
 \multicolumn{2}{l}{First-order }       &&&&&\\ 
       \multicolumn{2}{l}{interelectron}&$   502.083$&$   513.649$&$   537.458$&$   562.218$&                            TW\\ 
 \multicolumn{2}{l}{interaction}        &&&&&\\ 
 \multicolumn{2}{l}{Second-order}       &&&&&\\ 
       \multicolumn{2}{l}{interelectron}&$    -4.561$&$    -4.670$&$    -4.888$&$    -5.117$&                            TW\\ 
         \multicolumn{2}{l}{interaction}&$          $&$          $&$          $&$   -5.1140$&                 \cite{mohr00}\\ 
 \multicolumn{2}{l}{Third-order}        &&&&&\\ 
       \multicolumn{2}{l}{interelectron}&$     0.016$&$     0.017$&$     0.019$&$     0.020$&                            TW\\ 
 \multicolumn{2}{l}{interaction}        &&&&&\\ 
                            SE with NS &&$     0.615$&$     0.700$&$     0.899$&$     1.137$&          \cite{mohr92,mohr93}\\ 
                            VP with NS &&$    -0.173$&$    -0.193$&$    -0.239$&$    -0.297$&       \cite{johnson85,soff88}\\ 
                          SE screening &&$    -0.165$&$    -0.176$&$    -0.197$&$    -0.218$&          \cite{indelicato01}\\ 
                          VP screening &&$     0.047$&$     0.049$&$     0.060$&$     0.070$&       \cite{beier01artemyev}\\ 
                              Recoil   &&$     0.053$&$     0.054$&$     0.055$&$     0.057$&             \cite{artemyev95}\\ 
               \multicolumn{2}{l}{Total}&$-15023.203$&$-15535.132$&$-16591.468$&$-17692.311$&                            TW\\ 
 \end{tabular} 
 \label{tx5x3p0x2} 
 \end{table} 


\begin{table}
\caption{Different contributions to the total energy of the 
two-electron configuration 
$1s_{1/2}2p_{1/2} \,{}^3\! P_0$ (eV). 
The numbers in the Table present 
the ionization energy of the 
$2p_{1/2}$ 
electron with the opposite sign.} 
 \begin{tabular}{llrrrrc} 
 \multicolumn{2}{l}{Contribution}&$Z=\;$         80\phantom{88}& 
          91\phantom{88} 
 &         92\phantom{88}&         93\phantom{88} 
 &Ref.\\ 
 \hline 
          \multicolumn{2}{l}{Zero-order}&$-24622.160$&$-33320.132$&$-34215.481$&$-35130.460$&                           TW\\ 
   \multicolumn{2}{l}{Nuclear size (NS)}&$     0.75$&$     3.68$&$     4.41$&$     4.93$&                            TW\\ 
 \multicolumn{2}{l}{First-order }       &&&&&\\ 
       \multicolumn{2}{l}{interelectron}&$   702.993$&$   902.031$&$   923.176$&$   944.984$&                            TW\\ 
 \multicolumn{2}{l}{interaction}        &&&&&\\ 
 \multicolumn{2}{l}{Second-order}       &&&&&\\ 
       \multicolumn{2}{l}{interelectron}&$    -6.598$&$    -8.991$&$    -9.274$&$    -9.560$&                            TW\\ 
         \multicolumn{2}{l}{interaction}&$   -6.5959$&$          $&$   -9.2760$&$          $&                 \cite{mohr00}\\ 
 \multicolumn{2}{l}{Third-order}        &&&&&\\ 
       \multicolumn{2}{l}{interelectron}&$     0.034$&$     0.056$&$     0.058$&$     0.061$&                            TW\\ 
 \multicolumn{2}{l}{interaction}        &&&&&\\ 
                            SE with NS &&$     3.234$&$     8.772$&$     9.550$&$    10.376$&          \cite{mohr92,mohr93}\\ 
                            VP with NS &&$    -0.834$&$    -2.451$&$    -2.704$&$    -2.982$&       \cite{johnson85,soff88}\\ 
                          SE screening &&$    -0.416$&$    -0.857$&$    -0.897$&$    -0.937$&          \cite{indelicato01}\\ 
                          VP screening &&$     0.144$&$     0.338$&$     0.365$&$     0.391$&       \cite{beier01artemyev}\\ 
                              Recoil   &&$     0.068$&$     0.084$&$     0.086$&$     0.088$&             \cite{artemyev95}\\ 
               \multicolumn{2}{l}{Total}&$-23922.785$&$-32417.470$&$-33290.711$&$-34183.109$&                            TW\\ 
 \end{tabular} 
 \label{tx5x3p0x3} 
 \end{table} 


\begin{table} 
\caption{Different contributions to the total energy of the 
two-electron configuration 
$1s_{1/2}2s_{1/2} \,{}^3\! S_1$ (eV). 
The numbers in the Table present 
the ionization energy of the 
$2s_{1/2}$ 
electron with the opposite sign.} 
 \begin{tabular}{llrrrrc} 
 \multicolumn{2}{l}{Contribution}&$Z=\;$         60\phantom{88}& 
          62\phantom{88} 
 &         63\phantom{88}&         64\phantom{88} 
 &Ref.\\ 
 \hline 
          \multicolumn{2}{l}{Zero-order}&$-13063.004$&$-14015.106$&$-14506.589$&$-15008.567$&                            TW\\ 
   \multicolumn{2}{l}{Nuclear size (NS)}&$     0.928$&$     1.201$&$     1.342$&$     1.519$&                            TW\\ 
 \multicolumn{2}{l}{First-order }       &&&&&\\ 
       \multicolumn{2}{l}{interelectron}&$   333.849$&$   347.123$&$   353.854$&$   360.649$&                            TW\\ 
 \multicolumn{2}{l}{interaction}        &&&&&\\ 
 \multicolumn{2}{l}{Second-order}       &&&&&\\ 
       \multicolumn{2}{l}{interelectron}&$    -1.542$&$    -1.562$&$    -1.572$&$    -1.582$&                            TW\\ 
         \multicolumn{2}{l}{interaction}&$    -1.5455$&$          $&$          $&$          $&                 \cite{mohr00}\\ 
 \multicolumn{2}{l}{Third-order}        &$    -1.5459$&&&& \cite{asen02}\\
       \multicolumn{2}{l}{interelectron}&$    -0.001$&$    -0.001$&$    -0.001$&$    -0.001$&                            TW\\ 
 \multicolumn{2}{l}{interaction}        &&&&&\\ 
                            SE with NS &&$    11.409$&$    12.896$&$    13.700$&$    14.544$&          \cite{mohr92,mohr93}\\ 
                            VP with NS &&$    -1.621$&$    -1.893$&$    -2.042$&$    -2.201$&       \cite{johnson85,soff88}\\ 
                          SE screening &&$    -0.196$&$    -0.218$&$    -0.229$&$    -0.240$&          \cite{indelicato01}\\ 
                          VP screening &&$     0.052$&$     0.058$&$     0.061$&$     0.064$&       \cite{beier01artemyev}\\ 
                              Recoil   &&$     0.055$&$     0.057$&$     0.058$&$     0.059$&             \cite{artemyev95}\\ 
               \multicolumn{2}{l}{Total}&$-12710.071$&$-13657.445$&$-14141.418$&$-14635.756$&                            TW\\ 
 \end{tabular} 
 \label{tx5x3s1x1} 
 \end{table} 


\begin{table}
\caption{Different contributions to the total energy of the 
two-electron configuration 
$1s_{1/2}2s_{1/2} \,{}^3\! S_1$ (eV). 
The numbers in the Table present 
the ionization energy of the 
$2s_{1/2}$ 
electron with the opposite sign.} 
 \begin{tabular}{llrrrrc} 
 \multicolumn{2}{l}{Contribution}&$Z=\;$         65\phantom{88}& 
          66\phantom{88} 
 &         68\phantom{88}&         70\phantom{88} 
 &Ref.\\ 
 \hline 
          \multicolumn{2}{l}{Zero-order}&$-15521.201$&$-16044.661$&$-17124.764$&$-18250.361$&                            TW\\ 
   \multicolumn{2}{l}{Nuclear size (NS)}&$     1.696$&$     1.968$&$     2.386$&$     3.099$&                            TW\\ 
 \multicolumn{2}{l}{First-order }       &&&&&\\ 
       \multicolumn{2}{l}{interelectron}&$   367.512$&$   374.442$&$   388.524$&$   402.904$&                            TW\\ 
 \multicolumn{2}{l}{interaction}        &&&&&\\ 
 \multicolumn{2}{l}{Second-order}       &&&&&\\ 
       \multicolumn{2}{l}{interelectron}&$    -1.593$&$    -1.605$&$    -1.626$&$    -1.648$&                            TW\\ 
         \multicolumn{2}{l}{interaction}&$          $&$          $&$          $&$    -1.6548$&                 \cite{mohr00}\\ 
 \multicolumn{2}{l}{Third-order}        &&&&&\\ 
       \multicolumn{2}{l}{interelectron}&$     0.000$&$     0.000$&$     0.000$&$     0.000$&                            TW\\ 
 \multicolumn{2}{l}{interaction}        &&&&&\\ 
                            SE with NS &&$    15.427$&$    16.358$&$    18.363$&$    20.589$&          \cite{mohr92,mohr93}\\ 
                            VP with NS &&$    -2.374$&$    -2.557$&$    -2.962$&$    -3.419$&       \cite{johnson85,soff88}\\ 
                          SE screening &&$    -0.251$&$    -0.263$&$    -0.285$&$    -0.307$&          \cite{indelicato01}\\ 
                          VP screening &&$     0.067$&$     0.070$&$     0.081$&$     0.092$&       \cite{beier01artemyev}\\ 
                              Recoil   &&$     0.060$&$     0.061$&$     0.064$&$     0.067$&             \cite{artemyev95}\\ 
               \multicolumn{2}{l}{Total}&$-15140.657$&$-15656.187$&$-16720.219$&$-17828.984$&                            TW\\ 
 \end{tabular} 
 \label{tx5x3s1x2} 
 \end{table} 


\begin{table}
\caption{Different contributions to the total energy of the 
two-electron configuration 
$1s_{1/2}2s_{1/2} \,{}^3\! S_1$ (eV). 
The numbers in the Table present 
the ionization energy of the 
$2s_{1/2}$ 
electron with the opposite sign.} 
 \begin{tabular}{llrrrrc} 
 \multicolumn{2}{l}{Contribution}&$Z=\;$         80\phantom{88}& 
          91\phantom{88} 
 &         92\phantom{88}&         93\phantom{88} 
 &Ref.\\ 
 \hline 
          \multicolumn{2}{l}{Zero-order}&$-24622.160$&$-33320.132$&$-34215.481$&$-35130.460$&                            TW\\ 
   \multicolumn{2}{l}{Nuclear size (NS)}&$     9.34 $&$     32.50$&$       37.76$&$     40.94$&                            TW\\ 
 \multicolumn{2}{l}{First-order }       &&&&&\\ 
       \multicolumn{2}{l}{interelectron}&$     480.13$&$   578.374$&$     588.169$&$   598.188$&                            TW\\ 
 \multicolumn{2}{l}{interaction}        &&&&&\\ 
 \multicolumn{2}{l}{Second-order}       &&&&&\\ 
       \multicolumn{2}{l}{interelectron}&$    -1.789$&$    -1.995$&$      -2.018$&$    -2.040$&                            TW\\
\end{tabular} 
 \label{tx5x3s1x3} 
 \end{table} 


\begin{table}
\caption{Different contributions to the total energy of the 
three-electron configuration 
$(1s)^2 2s_{1/2}$ (eV). 
The numbers in the Table present 
the ionization energy of the 
$2s_{1/2}$ 
electron with the opposite sign.} 
 \begin{tabular}{llrrrrc} 
 \multicolumn{2}{l}{Contribution}&$Z=\;$         60\phantom{88}& 
          62\phantom{88} 
 &         63\phantom{88}&         64\phantom{88} 
 &Ref.\\ 
 \hline 
          \multicolumn{2}{l}{Zero-order}&$-13063.004$&$-14015.106$&$-14506.589$&$-15008.567$&                            TW\\ 
   \multicolumn{2}{l}{Nuclear size (NS)}&$     0.928$&$     1.201$&$     1.342$&$     1.519$&                            TW\\ 
 \multicolumn{2}{l}{First-order }       &&&&&\\ 
       \multicolumn{2}{l}{interelectron}&$   720.180$&$   749.835$&$   764.911$&$   780.160$&                            TW\\ 
 \multicolumn{2}{l}{interaction}        &&&&&\\ 
 \multicolumn{2}{l}{Second-order}       &&&&&\\ 
       \multicolumn{2}{l}{interelectron}&$    -8.954$&$    -9.124$&$    -9.213$&$    -9.303$&                            TW\\
\end{tabular} 
 \label{tx6xs12x1} 
 \end{table} 


\begin{table}
\caption{Different contributions to the total energy of the 
three-electron configuration 
$(1s)^2 2s_{1/2}$ (eV). 
The numbers in the Table present 
the ionization energy of the 
$2s_{1/2}$ 
electron with the opposite sign.} 
 \begin{tabular}{llrrrrc} 
 \multicolumn{2}{l}{Contribution}&$Z=\;$         65\phantom{88}& 
          66\phantom{88} 
 &         68\phantom{88}&         70\phantom{88} 
 &Ref.\\ 
 \hline 
          \multicolumn{2}{l}{Zero-order}&$-15521.201$&$-16044.661$&$-17124.764$&$-18250.361$&                            TW\\ 
   \multicolumn{2}{l}{Nuclear size (NS)}&$     1.696$&$     1.968$&$     2.386$&$     3.099$&                            TW\\ 
 \multicolumn{2}{l}{First-order }       &&&&&\\ 
       \multicolumn{2}{l}{interelectron}&$   795.587$&$   811.194$&$   842.996$&$   875.598$&                            TW\\ 
 \multicolumn{2}{l}{interaction}        &&&&&\\ 
 \multicolumn{2}{l}{Second-order}       &&&&&\\ 
       \multicolumn{2}{l}{interelectron}&$    -9.395$&$    -9.490$&$    -9.690$&$    -9.898$&                            TW\\
\end{tabular} 
 \label{tx6xs12x2} 
 \end{table} 


\begin{table}
\caption{Different contributions to the total energy of the 
three-electron configuration 
$(1s)^2 2s_{1/2}$ (eV). 
The numbers in the Table present 
the ionization energy of the 
$2s_{1/2}$ 
electron with the opposite sign.} 
 \begin{tabular}{llrrrrc} 
 \multicolumn{2}{l}{Contribution}&$Z=\;$         80\phantom{88}& 
          91\phantom{88} 
 &         92\phantom{88}&         93\phantom{88} 
 &Ref.\\ 
 \hline 
          \multicolumn{2}{l}{Zero-order}&$-24622.160$&$-33320.132$&$-34215.481$&$-35130.460$&                            TW\\ 
   \multicolumn{2}{l}{Nuclear size (NS)}&$     9.34$&$    32.50$&$    37.76$&$    40.94$&                            TW\\ 
 \multicolumn{2}{l}{First-order }       &&&&&\\ 
       \multicolumn{2}{l}{interelectron}&$   795.587$&$   811.194$&$   842.996$&$   875.598$&                            TW\\ 
 \multicolumn{2}{l}{interaction}        &&&&&\\ 
 \multicolumn{2}{l}{Second-order}       &&&&&\\ 
       \multicolumn{2}{l}{interelectron}&$    -11.143$&$  -13.028$&$   -13.228$&$   -13.441$&                            TW\\
\end{tabular} 
 \label{tx6xs12x3} 
 \end{table} 


\begin{table}
\caption{Different contributions to the total energy of the 
three-electron configuration 
$(1s)^2 2p_{1/2}$ (eV). 
The numbers in the Table present 
the ionization energy of the 
$2p_{1/2}$ 
electron with the opposite sign.} 
 \begin{tabular}{llrrrrc} 
 \multicolumn{2}{l}{Contribution}&$Z=\;$         60\phantom{88}& 
          62\phantom{88} 
 &         63\phantom{88}&         64\phantom{88} 
 &Ref.\\ 
 \hline 
          \multicolumn{2}{l}{Zero-order}&$-13063.004$&$-14015.106$&$-14506.589$&$-15008.567$&                            TW\\ 
   \multicolumn{2}{l}{Nuclear size (NS)}&$     0.038$&$     0.052$&$     0.061$&$     0.071$&                            TW\\ 
 \multicolumn{2}{l}{First-order }       &&&&&\\ 
       \multicolumn{2}{l}{interelectron}&$   875.619$&$   913.798$&$   933.291$&$   953.066$&                            TW\\ 
 \multicolumn{2}{l}{interaction}        &&&&&\\ 
 \multicolumn{2}{l}{Second-order}       &&&&&\\ 
       \multicolumn{2}{l}{interelectron}&$   -15.092$&$   -15.506$&$   -15.726$&$   -15.951$&                            TW\\
\end{tabular} 
 \label{tx6xp12x1} 
 \end{table} 


\begin{table}
\caption{Different contributions to the total energy of the 
three-electron configuration 
$(1s)^2 2p_{1/2}$ (eV). 
The numbers in the Table present 
the ionization energy of the 
$2p_{1/2}$ 
electron with the opposite sign.} 
 \begin{tabular}{llrrrrc} 
 \multicolumn{2}{l}{Contribution}&$Z=\;$         65\phantom{88}& 
          66\phantom{88} 
 &         68\phantom{88}&         70\phantom{88} 
 &Ref.\\ 
 \hline 
          \multicolumn{2}{l}{Zero-order}&$-15521.201$&$-16044.661$&$-17124.764$&$-18250.361$&                            TW\\ 
   \multicolumn{2}{l}{Nuclear size (NS)}&$     0.083$&$     0.099$&$     0.129$&$     0.180$&                            TW\\ 
 \multicolumn{2}{l}{First-order }       &&&&&\\ 
       \multicolumn{2}{l}{interelectron}&$   973.133$&$   993.500$&$  1035.185$&$  1078.206$&                            TW\\ 
 \multicolumn{2}{l}{interaction}        &&&&&\\ 
 \multicolumn{2}{l}{Second-order}       &&&&&\\ 
       \multicolumn{2}{l}{interelectron}&$   -16.182$&$   -16.422$&$   -16.919$&$   -17.450$&                            TW\\
\end{tabular} 
 \label{tx6xp12x2} 
 \end{table} 


\begin{table}
\caption{Different contributions to the total energy of the 
three-electron configuration 
$(1s)^2 2p_{1/2}$ (eV). 
The numbers in the Table present 
the ionization energy of the 
$2p_{1/2}$ 
electron with the opposite sign.} 
 \begin{tabular}{llrrrrc} 
 \multicolumn{2}{l}{Contribution}&$Z=\;$         80\phantom{88}& 
          91\phantom{88} 
 &         92\phantom{88}&         93\phantom{88} 
 &Ref.\\ 
 \hline 
          \multicolumn{2}{l}{Zero-order}&$-24622.160$&$-33320.132$&$-34215.481$&$-35130.460$&                            TW\\ 
   \multicolumn{2}{l}{Nuclear size (NS)}&$       0.75$&$       3.68$&$       4.41$&$       4.93$&                            TW\\ 
 \multicolumn{2}{l}{First-order }       &&&&&\\ 
       \multicolumn{2}{l}{interelectron}&$   1317.19$&$   1642.274$&$  1676.142$&$  1710.926$&                            TW\\ 
 \multicolumn{2}{l}{interaction}        &&&&&\\ 
 \multicolumn{2}{l}{Second-order}       &&&&&\\ 
       \multicolumn{2}{l}{interelectron}&$   -20.827$&$   -26.027$&$   -26.597$&$   -27.232$&                            TW\\
\end{tabular} 
 \label{tx6xp12x3} 
 \end{table} 


\begin{table}
 \caption{Different theoretical data for 
the energy levels of two-electron configurations. 
The numbers in the Table present 
the ionization energy of the 
$2s_{1/2}$ 
or 
$2p_{1/2}$ 
electron with the opposite sign, respectively.} 
 \begin{tabular}{lrrrrrrrr} 
 Contribution&$Z=\;$60\phantom{88}&62\phantom{88} 
 &63\phantom{88}&64\phantom{88}&65\phantom{88} 
 &66\phantom{88}&68\phantom{88}&70\phantom{88}\\ 
 \hline 
 $E(2 {}^1\! S_{0})$\\ 
                         \quad This work&$-12618.333$&$-13549.608$&$-14030.409$&$-14521.491$&$-15023.050$&$-15535.148$&$-16592.030$&$-17693.252$\\ 
              \quad Drake \cite{drake88}&$-12618.188$&$-13549.412$&$-14030.191$&$-14521.251$&$-15022.780$&$-15534.938$&$-16591.716$&$-17692.948$\\ 
      \quad Plante et al \cite{plante94}&$-12618.629$&$-13549.909$&$          $&$-14521.817$&$          $&$-15535.580$&$-16592.446$&$-17693.778$\\ 
 \hline 
 $E(2 {}^3\! P_{0})$\\ 
                         \quad This work&$-12619.519$&$-13550.416$&$-14031.002$&$-14521.881$&$-15023.203$&$-15535.132$&$-16591.468$&$-17692.311$\\ 
              \quad Drake \cite{drake88}&$-12619.024$&$-13549.822$&$-14030.359$&$-14521.184$&$-15022.452$&$-15534.324$&$-16590.550$&$-17691.276$\\
\end{tabular} 
 \label{tx7x1} 
 \end{table} 


\begin{table}
 \caption{Different theoretical data for 
the energy levels of two-electron configurations. 
The numbers in the Table present 
the ionization energy of the 
$2s_{1/2}$ 
or 
$2p_{1/2}$ 
electron with the opposite sign, respectively.} 
 \begin{tabular}{lrrrrrrrr} 
 Contribution&$Z=\;$80\phantom{88}&91\phantom{88} 
 &92\phantom{88}&93\phantom{88}\\ 
 \hline 
 $E(2 {}^1\! S_{0})$\\ 
                         \quad This work&$ -23925.626$&$ -32415.499$&$ -33286.200$&$ -34178.051$\\ 
              \quad Drake \cite{drake88}&$ -23924.725$&$ -32413.255$&$ -33284.719$&$ -34175.303$\\
      \quad Plante et al \cite{plante94}&$ -23926.313$&$ -13549.909$&$ -33288.445$&$           $\\ 
 \hline 
 $E(2 {}^3\! P_{0})$\\ 
                         \quad This work&$ -23922.785$&$ -32417.470$&$ -33290.711$&$ -34183.109$\\ 
              \quad Drake \cite{drake88}&$ -23920.774$&$ -32413.464$&$ -33286.535$&$ -34178.540$\\
\end{tabular} 
 \label{tx7x2} 
 \end{table} 
%
%
%
%
%
%

%
%
%
%
%
%
\end{document}